\documentclass[a4paper]{article}
\usepackage{colortbl}
\usepackage[svgnames]{xcolor}
\usepackage{fullpage} 
\usepackage{amssymb}
\usepackage{amsthm}
\usepackage{amsmath} 
\usepackage{amsfonts}
\usepackage{epstopdf}
\usepackage[version=3]{mhchem}
\usepackage{multicol}
\usepackage{hyperref}
\usepackage{graphicx}
\usepackage{color}
\usepackage{soul}
\usepackage[margin=1.5cm,top=2.5cm,bottom=2.8cm,centering]{geometry}
\usepackage[font=small,labelfont=bf]{caption}
\usepackage{setspace}
\usepackage{rotating}

 
\begin{document}

\title{\begin{Large}Sound and light Doppler effects\end{Large}}

\author{Denis Michel \ \href{https://orcid.org/0000-0002-0643-8025}{\includegraphics[scale=0.05]{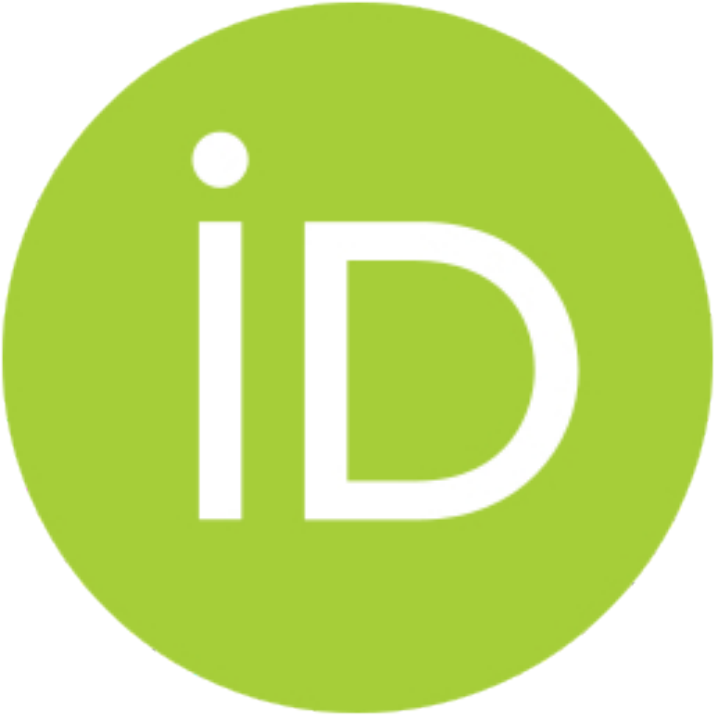}}}
\date{\begin{small} Université de Rennes1, Irset, Rennes, France. E-mail: denis.michel@live.fr \end{small}}
\maketitle
\begin{multicols}{2}
\noindent
\textbf{Although electromagnetic and acoustic waves profoundly differ in their nature, comparing their Doppler effects is instructive and reveals persistent conceptual traps. The principle of the Doppler effect was presented by Christian Doppler in 1842 long before the advent of relativity theory, but while the relativistic Doppler effect is now well established, its non-relativistic version retrospectively called classical, suffers from misleading intuitions such as: (1) there is no transverse classical Doppler effect; (2) the Galilean Doppler effect corresponds to the relativistic one without Lorentz dilation factor. Additional pitfalls concern the sound Doppler effect which, in addition to being Galilean, results from asymmetric contributions of the sources and receptors and depends on a material medium for its propagation, possibly modifying the effective velocity of the wave. Moreover, contrary to light, the information on the sound Doppler effect and the location of the source are transmitted through different channels, sound and light respectively, thereby complicating data interpretation and increasing confusion. A thorough revision is proposed here addressing these issues and providing a complete set of new candidate Doppler formulas, non-collinear, two- and three-dimensional and in their angular and linear versions. \\}


\noindent
\textbf{Highlights:}\\ \begin{small}
$ \bullet $ Two- and three-dimensional angular Doppler formulas are reconstructed. \\
$ \bullet $ The Galilean transverse Doppler effect exists. \\
$ \bullet $ It is technically unfeasible to verify the relativistic transverse Doppler effect of Einstein. \\
$ \bullet $ The traditional formula for non-collinear sound Doppler effect is disqualified.\\
$ \bullet $ A general candidate Doppler formula is proposed for the sound Doppler effect in the presence of wind. \\ 
$ \bullet $ Longitudinal non-collinear and non-angular Doppler formulas are presented.\\ 
\end{small}
\vspace{-0.5cm}
\section{Introduction}

One of the major results of the theory of the special relativity theory was the establishment of the laws of light aberration and Doppler effect, but the Doppler effect named for its inventor, extends beyond electromagnetic waves and also exists for waves obeying Galilean rules. Although relativistic transformations are more subtle than Galilean transformations, Galilean Doppler effects appear more complicated and, as a matter of fact, their current formulations are misled. According to the historical reminders of \cite{Nolte2020}, one of the difficulties encountered by Christian Doppler was that his theory seemed too simple mathematically to describe physics, at the time when the most celebrated tool was the differential equations. However, if some differential systems present technical subtleties, they remain globally intuitive while conversely, the conception of mathematically simple processes may be tricky, as is typically illustrated by the classical Doppler effect. The first sections of this study will focus on the Doppler effect understood as Galilean in the mathematical sense, by considering the receiver as immobile. Then the additional particularities of the sound Doppler will be introduced, including the source-receiver asymmetry and the displacement of the propagation medium.

\section{Problems with the angular version of the classical Doppler effect}

\subsection{The currently accepted formula}

The current Doppler formula describing the change in frequency of a moving source perceived by a static observer is

\begin{equation} \dfrac{f^{\textup{mov}}}{f}  = \dfrac{1}{1-\beta \cos \theta} \end{equation} 

where $ \beta $ is the ratio of the source velocity to the wave velocity ($ \beta =v/c $). A first ambiguity in this formula concerns the nature of the angle $ \theta $ between the trajectory of the source and the direction of the receiver. The phenomenon of aberration is not specific to relativity but it is strangely neglected in the so called classical Doppler effect. The origin of the angle considered is générally the source but in addition it is necessary to specify whether the position of the source is to be considered when the wave is emitted or received. As the Doppler effect is naturally carried and detected through the same wave, the first possibility appears reasonable and it would therefore be the angle $ \theta ' $ of Fig.1, 
\begin{equation} \dfrac{f^{\textup{mov}}}{f} = \dfrac{1}{1-\beta \cos \theta ' }\label{Eq:dop-trad} \end{equation}
even if this angle is considered as apparent by the observer who knows that it no longer corresponds to the true position of the source when the Doppler effect is received. \\

\label{fig:sound-light}
\begin{center}
\includegraphics[width=6.5cm]{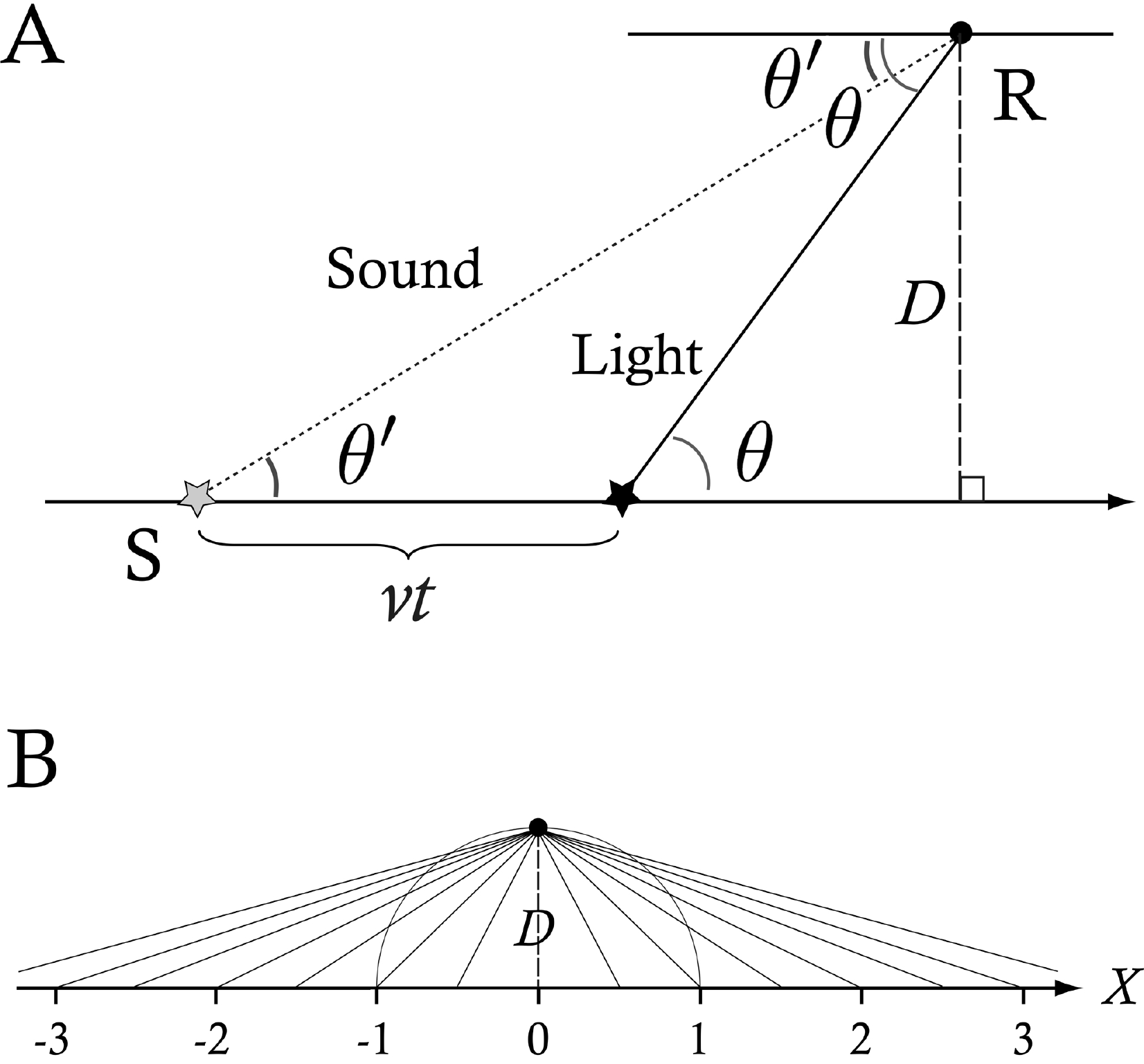}
\end{center}
\begin{small}\textbf{Figure 1}. Parameters used to represent the sound Doppler effect. (\textbf{A}) Angular representation. As the speed of light can be considered as infinite compared to that of sound, the sound wave carrying the Doppler effect emitted by the source $ S $ under an angle $ \theta' $ towards the receiver $ R $, and the light carrying the image of the source, emitted under an angle $ \theta $ towards the receiver, arrive together at the receiver. On the whole trajectory, $ D $ is the shortest distance separating the source and the receiver. (\textbf{B}) In this linear representation, the $ X $ axis corresponds to the trajectory of the source. The unit of length chosen here is the shortest distance $ D $, which allows to define the normalized distance $ x=X/D $.  \\ \end{small} 

\noindent
Eq.(\ref{Eq:dop-trad}) is given in textbooks \cite{HRW} and online courses \cite{Fowler}, but disagrees with the hypothesis of a spherical propagation of the wave around its point of emission, which predicts a Doppler formula of the form \cite{dmichel2022} 

\begin{equation}  \dfrac{f^{\textup{mov}}}{f} = \dfrac{1}{ \sqrt{1+\beta^{2}-2 \beta \cos \theta'} } \label{eq:dop-gal} \end{equation}

In an attempt to reconcile the traditional (Eq.(\ref{Eq:dop-trad})) and exact (Eq.(\ref{eq:dop-gal})) formulas, the former could be rewritten as a square root of a square,

\begin{equation} \dfrac{f^{\textup{mov}}}{f} = \dfrac{1}{\sqrt{1+\beta^{2}\cos^{2} \theta' -2 \beta \cos \theta'} } \label{eq:dop-err}  \end{equation}

But Eq.(\ref{eq:dop-gal}) and Eq.(\ref{eq:dop-err}) are still different and in addition one cannot invoke the approximation of a very small angle $ \theta' $ (which would make $ \cos^{2} \theta' \approx 1 $) because of the alleged prediction that the Doppler effect must vanish ($ \lambda ^{\textup{mov}}/\lambda = 1 $) when $ \theta' =\pi/2 $ and $ \cos \theta' =0 $. So we see that the classical Doppler formula is based on several layers of misleading intuitions. To make matters worse, these errors are likely to have been supported by a resemblance with the relativistic Doppler effect.

\subsection{How the relativistic Doppler effect could have reinforced a flawed classical approach}
The formula of the classical Doppler effect (Eq.(\ref{Eq:dop-trad})), has likely been consolidated by the advent of the relativistic Doppler effect, whose form can suggest the confusing idea that it is simply the classical Doppler effect corrected by the relativistic dilation factor. 

\subsubsection{The puzzling idea that the classical Doppler effect is the primary effect of the relativistic one}

According to approximate relativistic theories, the classical formula would be a "primary" Doppler effect of purely kinetic nature, which must be completed with a so-called "secondary" effect of time dilation by the Lorentz factor (multiplication of the periods by $ 1/\sqrt{1-\beta^{2}} $) to obtain the relativistic Doppler effect. In fact, in relativity the kinetic and temporal effects cannot be dissociated and the Lorentz factor itself includes the change of kinetic energy. Nevertheless, this questionable principle was accepted because it apparently works. In the famous longitudinal relativistic Doppler formulas, for the approach,

\begin{subequations} \label{eq:collinear}
\begin{equation} \dfrac{f^{\textup{mov}}}{f} =\dfrac{\sqrt{1-\beta^{2}}}{1-\beta} = \sqrt{\dfrac{1+\beta}{1-\beta}} \end{equation} 
and for the recession,
\begin{equation} \dfrac{f^{\textup{mov}}}{f} =\dfrac{\sqrt{1-\beta^{2}}}{1+\beta} = \sqrt{\dfrac{1-\beta}{1+\beta}} \end{equation} 

\noindent
It can be extended to all intermediate points, not collinear to the source path, by applying the same correction to the general equation Eq.(\ref{Eq:dop-trad}).

\begin{equation} \dfrac{f^{\textup{mov}}}{f} =\dfrac{\sqrt{1-\beta ^{2}}}{1-\beta \ \cos \theta '}  \label{eq:TRD} \end{equation} 
\end{subequations}

It happens that this latter equation is indeed Einstein's relativistic Doppler formula where $ \theta' $ is the reception angle. This unfortunate identity has logically reinforced the supposed validity of Eq.(\ref{Eq:dop-trad}) as the classical Doppler effect formula for generations of researchers and teachers, and convinced them that the relativistic Doppler effect of Eq.(\ref{eq:TRD}) is simply the classical Doppler effect of Eq.(\ref{Eq:dop-trad}) corrected by the dilatation factor of special relativity. The consensus generated by this apparent evidence likely inhibited naive questions such as, for instance, if the only difference between the classical and relativistic Doppler effects is the dilation of the periods for the latter, then why would the classical and relativistic aberration rules be different? Indeed, the generalized correction by the dilation factor has a homothetic effect unable in itself to modify the angles. But a majority consensus in the scientific community naturally tends to inhibit legitimate questions. \\
In addition, the secondary Doppler effect supposedly specific to relativity, was further consolidated by the inappropriate use of the arithmetic mean in the most famous validation tests of the relativistic Doppler effect, either longitudinal \cite{Ives} or transversal \cite{Hasselkamp} (appendix B). As we will see in section 6, the idea that this relativistic secondary Doppler effect is the only one perceptible in the transverse situation, was based on a misconception of the Galilean Doppler effect which presents in fact a very similar transverse Doppler effect.

\section{Galilean and relativistic wave bubble equations arising from galilean and relativistic transformations}

Galilean Doppler and aberration formulas can be deduced from the rays connecting the source to the wavefront, taking into account that the wavefront of a classical wave is the surface of a sphere and that of a relativistic wave is the surface of an ellipsoid with rotational symmetry around the axis of the trajectory \cite{Poincare1918}. \\

\subsection{The Galilean wave bubble}

The Galilean transformations are simple:
 $ x'=x+vt, \ y'=y, \ z'=z $ and $ t'=t $. Hence, the surface of the Galilean wavefront emitted from a source moving at speed $ v $ in the $ x $ direction, is a sphere of Cartesian equation
 
\begin{subequations} 
\begin{equation} (x+vt)^{2}+y^{2}+z^{2}=(ct)^2 \end{equation}
For a single period wave bubble ($ t=T=1 $) and using the parameter $ \beta=v/c $
\begin{equation} (x+\beta)^{2}+y^{2}+z^{2}=1 \label{Eq:3Dsphere} \end{equation}
\end{subequations} 

The wavelengths are directly obtained by converting this Cartesian equation in polar equation \cite{dmichel2022}, as the radii $ \rho $ using polar and deviation angles such that $ x=\rho \sin \theta \cos \varphi $, $ y=\rho \sin \theta \sin \varphi $ and $ z=\rho \cos \theta $. The radius of the Galilean wave bubble derived from the complete Cartesian equation Eq.(\ref{Eq:3Dsphere}) reads
\begin{equation} \rho= \sqrt{1-\beta^{2}(1-\sin^{2} \theta \cos^{2} \varphi})-\beta \sin \theta \cos \varphi \label{Eq:3D-spher} \end{equation}
\noindent
shown in Fig.2A and which reduces in the two spatial dimensions $ x=\rho \cos \theta $ and $ y=\rho \sin \theta $, to the Galilean circle

\begin{equation} \rho = \dfrac{\lambda^{\textup{mov}}}{\lambda}=\sqrt{1-\beta^{2} \sin^{2}  \theta} - \beta \cos  \theta \label{Eq:lambda-gal} \end{equation} 

which is equivalent to a Doppler effect formula expressed using wavelentghs. However, this Doppler effect is not valid in optics since the Doppler effect of electromagnetic waves, called relativistic, should take into account the relativistic transformations.

\subsection{The relativistic wave bubble}
Lorentz transformations rewritten by Poincar\'e \cite{Poincare1905} assuming $ c=1 $, are:
\noindent
$ x'=\dfrac{x+\beta t}{\sqrt{1-\beta^{2}}}, \ y'=y, \ z'=z $ and $ t'=\dfrac{t+\beta x}{\sqrt{1-\beta^{2}}} $. They transform the coordinate $ x $ into $ x'=x\sqrt{1-\beta^{2}} + \beta t' $, so that the relativistic wavefront is, for one period $ t'=T=1 $ and the wavelength $ cT=1 $, the surface of the single-period ellipsoid:
 
\begin{equation} (x\sqrt{1-\beta^{2}}+\beta)^{2}+y^{2}+z^{2}=1 \label{Eq:cart-relat} \end{equation} 
\noindent

which gives upon conversion in polar coordinates the remarkably elegant equation

\begin{equation} \rho = \dfrac{\sqrt{1-\beta^{2}}}{1+\beta \sin \theta \cos  \varphi} \label{Eq:3D-relat} \end{equation}
\noindent
shown in Fig.2B and reducing in 2D polar coordinates to an ellipse, better known in normal coordinates as the relativistic Doppler effect

\begin{equation} \rho = \dfrac{\lambda^{\textup{mov}}}{\lambda}= \dfrac{\sqrt{1-\beta^{2}}}{1+\beta \cos  \theta} \label{Eq:lambda-relat} \end{equation}

\begin{center}
\includegraphics[width=8.8cm]{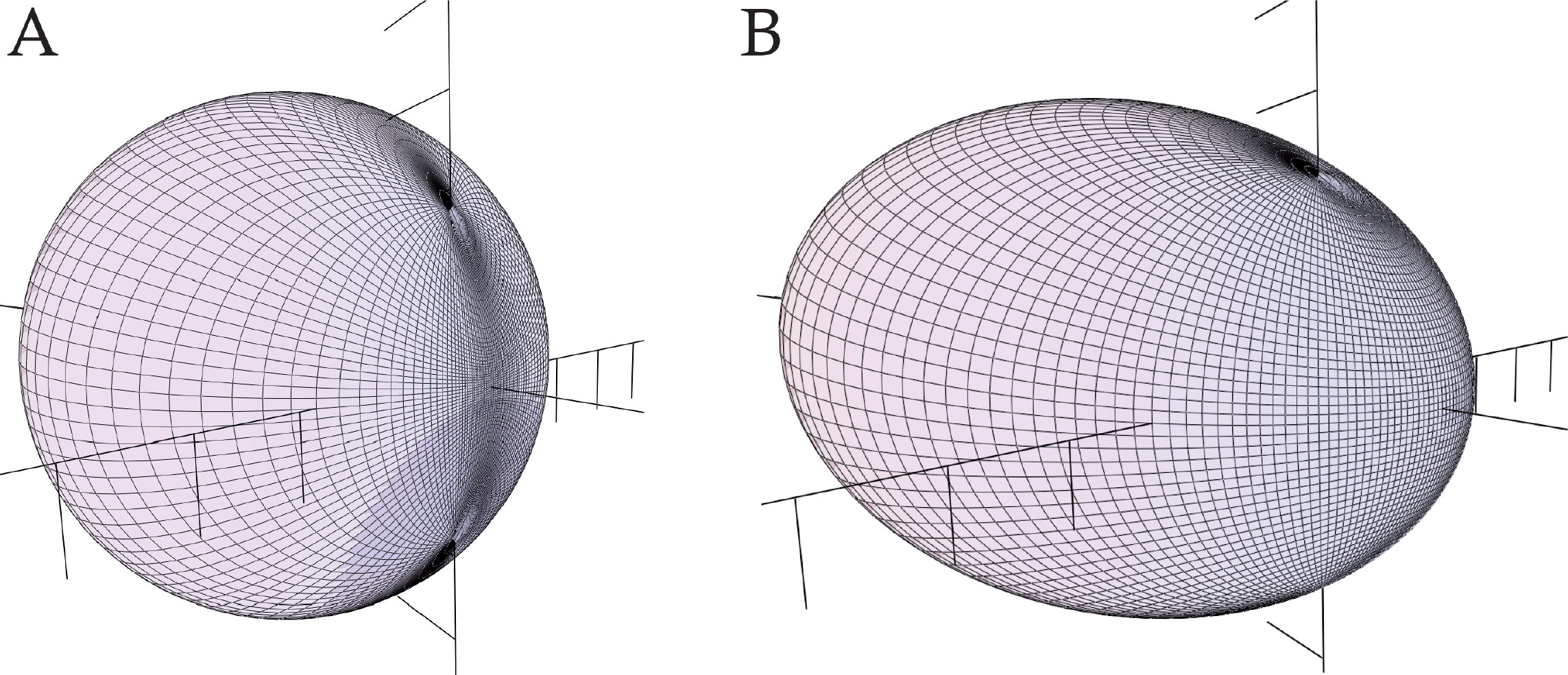}\label{fig:paramspheric}
\end{center}
\begin{small}\textbf{Figure 2}. Perspective view of the shapes of the Galilean (\textbf{A}) and relativistic (\textbf{B}) single period wavefront surfaces, drawn to Eq.(\ref{Eq:3D-spher}) and Eq.(\ref{Eq:3D-relat}) respectively, for $ \beta =0.9 $. The source is located at the intersection of the axes. \end{small} \\

\section{Galilean and relativistic Doppler and aberration formulas}

The above equations of wave bubbles are also those of Doppler effects expressed in wavelengths. They can be more appropriately expressed with frequencies, in the Galilean case for an immobile receptor in a stationary propagation medium. The angle $ \theta $ used in these equations takes into account the position of the source when the Doppler effect is received, but as this position is not visible by the observer in the case of light, it is useful to define another angle, $ \theta' $ taking into account the former position of the source precisely when it emitted the wave whose Doppler effect is detected.

\subsection{Two usable angles and two sets of formulas}
\noindent
Two angles, linked together by aberration relations, can be used to describe Doppler effects:\\
\noindent
$ \bullet $ the angle $ \theta $, defined between the source path and the direction of the source-receiver line when the Doppler effect is received, and\\
\noindent
$ \bullet $ the angle $ \theta' $, defined between the trajectory of the source and the line connecting the point of emission to the receiver (Fig.1A).

\subsection{The Galilean formulas}

The reciprocal relations of aberration linking these two angles are
\begin{subequations}  
\begin{equation} \cos \theta = \dfrac{\cos  \theta' - \beta}{\sqrt{1+\beta^{2}-2\beta \cos  \theta' }} \label{Eq:s-aber} \end{equation}
and
\begin{equation} \cos \theta' = \cos \theta \ \sqrt{1-\beta ^{2} \sin^{2} \theta} + \beta \sin^{2} \theta  \label{Eq:s-aber'} \end{equation}
and using the tangent function often employed to describe the aberration,
\begin{equation}  \tan \theta'=    \dfrac{\sin \theta \left(\sqrt{1-\beta^{2} \sin^{2} \theta} -\beta \cos \theta \right)}{ \cos \theta \sqrt{1-\beta^{2} \sin^{2} \theta}+ \beta \sin^{2} \theta } \end{equation}
\end{subequations}   

\noindent
Accordingly, two Doppler formulas related through the aberration relations, can be established depending on the angle considered. The Doppler effect is, in function of $ \theta $,
\begin{subequations}  
\begin{equation} \left(\dfrac{f^{\textup{mov}}}{f}\right)_{\theta}= \dfrac{1}{\sqrt{1-\beta^{2} \sin^{2}  \theta} - \beta \cos  \theta}  \label{eq:s-D(theta)} \end{equation}
and in function of $ \theta' $
\begin{equation} \left(\dfrac{f^{\textup{mov}}}{f}\right)_{\theta'} = \dfrac{1}{\sqrt{1+\beta^{2}-2 \beta \cos  \theta'}} \label{eq:s-D(theta')} \end{equation}
\end{subequations}   

\noindent
corresponding to Eq.(\ref{eq:dop-gal}).

\subsection{The relativistic formulas}

The above formulas are Galilean and do not apply to electromagnetic waves whose wavefronts are ellipsoidal surfaces \cite{Poincare1918,Einstein}. The aberration relations are in this case

\begin{subequations}  
\begin{equation} \cos \theta = \dfrac{\cos  \theta' - \beta}{1-\beta \cos  \theta'}  \label{Eq:e_aber} \end{equation} 
\begin{equation} \cos \theta' =  \dfrac{\cos  \theta + \beta}{1+\beta \cos  \theta}  \label{Eq:e-aber'} \end{equation} 
\begin{equation} \tan \theta'=  \dfrac{\sin \theta \sqrt{1-\beta^{2}}}{\beta + \cos \theta}  \end{equation}
\end{subequations} 
\noindent
and the Doppler effect can also be written in two ways
\begin{subequations}  
\begin{equation} \left(\dfrac{f^{\textup{mov}}}{f}\right)_{\theta} =  \dfrac{1+\beta \cos  \theta}{\sqrt{1-\beta^{2}}} \label{Eq:e-D(theta)} \end{equation}
\begin{equation} \left(\dfrac{f^{\textup{mov}}}{f}\right)_{\theta'} =  \dfrac{\sqrt{1-\beta^{2}}}{1-\beta \cos  \theta'} \label{Eq:e-D(theta')} \end{equation}
\end{subequations}  

The relativistic formulas are demonstrated by the standard relativistic approach in \cite{Einstein} further detailed in \cite{Joos}. For alternative geometric demonstrations of all these formulas, Galilean and relativistic, see \cite{dmichel2022}. Although the Galilean transformations are much simpler than the relativistic (or Lorentz) ones, paradoxically the relativistic Doppler and aberration formulas appear simpler and more elegant than their Galilean counterparts, even before applying them additional complications detailed later, including the propagation medium and the source-receptor asymmetry. The Doppler profiles described by these equations are shown in Fig.3 and Fig.4 as functions of $ \theta $ and $ \theta' $. 
\label{fig:galilean}
\begin{center}
\includegraphics[width=8cm]{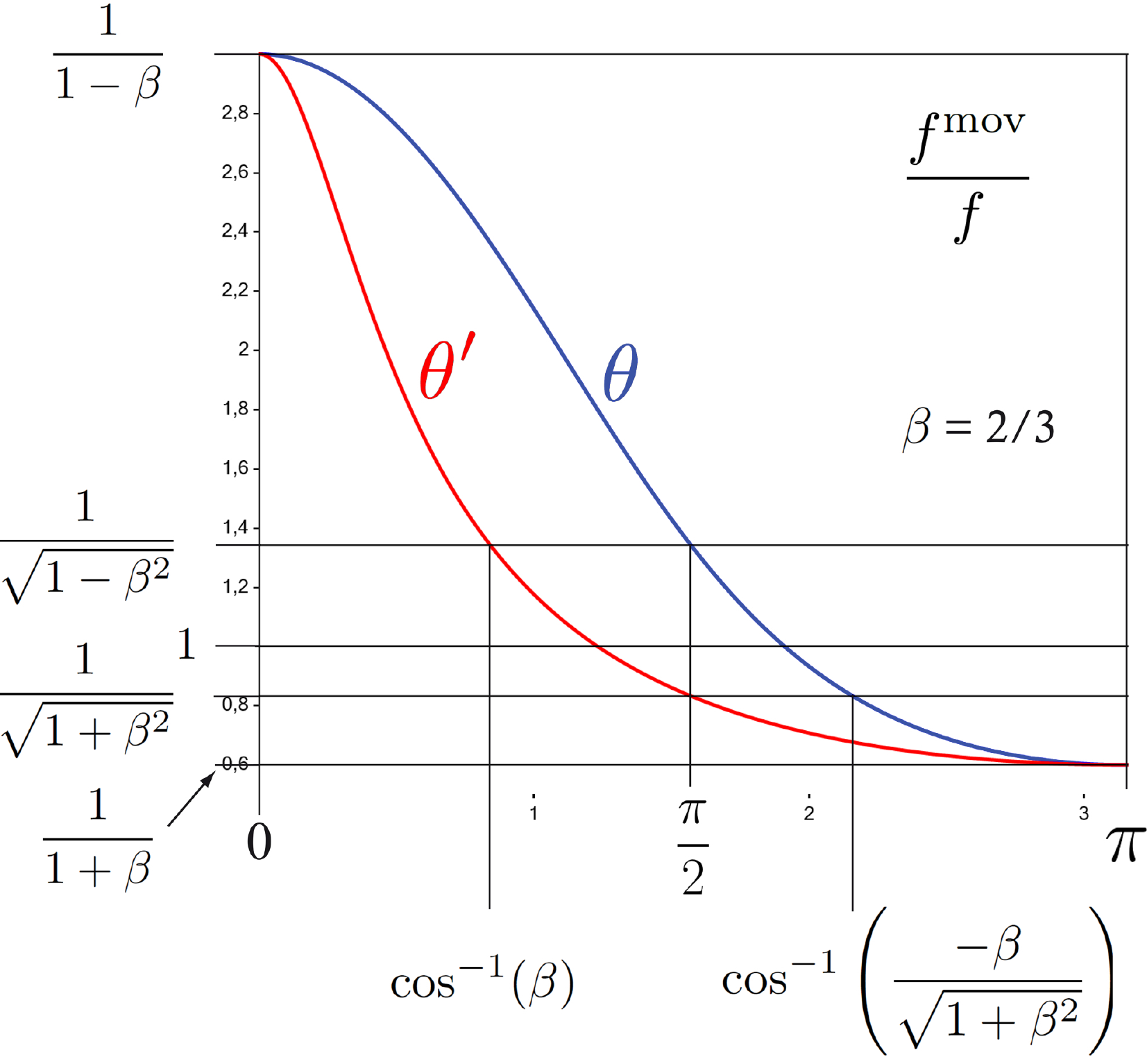}
\end{center}
\begin{small}\textbf{Figure 3}. Angular representation of the Doppler effect of a Galilean wave emitted  towards a fixed receiver by a moving source. \\ \end{small} 

$ \theta' $ refers to the real point of emission of the wave and $ \theta $ refers to the real position of the source, the difference between the two being due to the delay in the travel of the wave whose speed is finite. This comparison of the Galilean (Fig.3) and relativistic (Fig.4) profiles clearly shows that: (i) The only common point between them, giving the same Doppler effect, is $ (\theta, \theta') =(\pi/2, \cos^{-1} \beta) $. (ii) The difference between the Galilean and relativistic Doppler effects is not due to the absence of a transverse effect for the first one contrary to the second one, because the ratio $ f^{\textup{mov}}/f =1 $ is never obtained for an angle of $ \pi/2 $. (iii) Contrary to popular belief, the relativistic Doppler effect is not merely the Galilean Doppler effect corrected by the Lorentz expansion factor, as erroneously suggested by the usual Eq.(\ref{Eq:dop-trad}) supposed to be the relativistic formula Eq.(\ref{Eq:e-D(theta')}) lacking the Lorentz factor. As a matter of fact, the polar curve described by Eq.(\ref{Eq:dop-trad}) is an ellipse but not a circle, which definitely disqualifies it as a Galilean Doppler formula.\\

\label{fig:relativistic}
\begin{center}
\includegraphics[width=7.7cm]{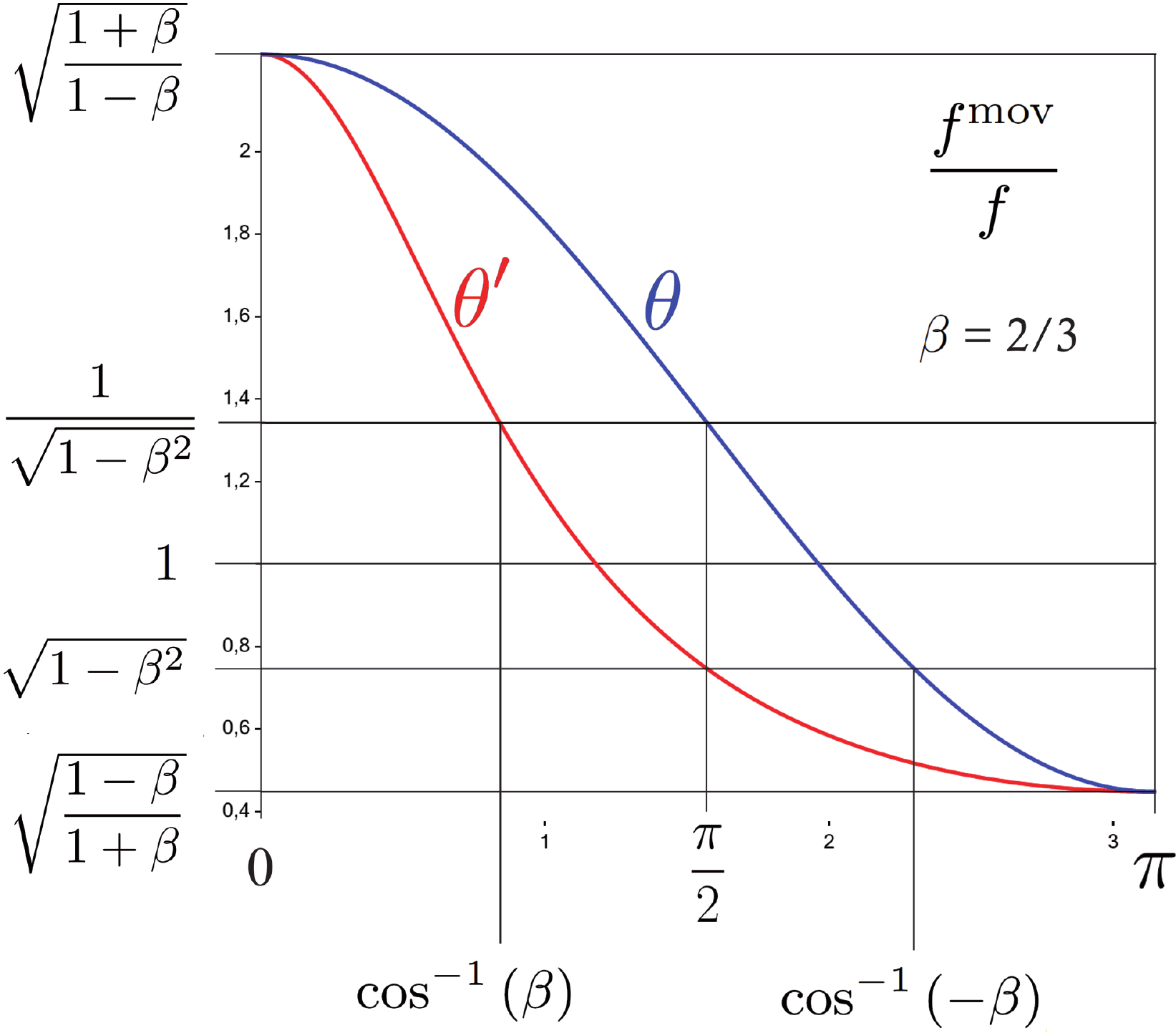}
\end{center}
\begin{small}\textbf{Figure 4}. Angular representation of the Doppler effect of electromagnetic waves, called relativistic.  \\ \end{small} 

\subsection{Dual information on the Doppler effect and the position of the source}

Since no information can exceed the speed of light, relativistic formulas do not suffer from any ambiguity in their measurement, because the information on the position of the source and on the Doppler effect necessarily pass through the same channel. This is not the case for the Doppler effect of low speed waves because the actual position of the source can be determined visually, i.e. transmitted by light \cite{Mangiarotty}. Therefore, the information on the sound and the position of the source, arriving through different channels (Fig.1A) can be combined (as illustrated in the appendix D).\\

\section{Doppler effects as functions of distance}

\subsection{Converting angles to distances}

The representations of the Doppler effect as a function of the angle of the source varying from 0 to $ \pi $ shown in Figs.3 and 4, amplify the central part of the trajectory around $ \pi/2 $, and compress the more distant locations until infinity. Hence, they are poorly appropriate to ordinary measurements for a source evolving at constant speed along a linear coordinate $ X $, as that shown in the appendix D. To superimpose these records on a theoretical curve, it is first necessary to establish the correspondence between angles and distances. Several units can be chosen for the $ X $ axis, one of the simplest consists in giving to the increment of $ X $ the value of the minimum distance $ D $ between the source a the receiver. We can then define a relative distance $ x=X/D $ (Fig.1B) which allows to standardize all the data for sources passing more or less far away. The correspondence of any angle $ \vartheta $ with the distance is
\begin{equation} D = - X \tan \vartheta \end{equation} 

\begin{equation} \dfrac{X}{D}=x= - \dfrac{\cos \vartheta}{\sqrt{1-\cos^{2} \vartheta}} \end{equation} 

\begin{equation} \cos \vartheta = - \dfrac{x}{\sqrt{1+x^{2}}} \label{Eq:cos-x} \end{equation} 

By applying this relation to the angles $ \theta $ and $ \theta' $ presented previously, we obtain, for the classical Doppler formula

\begin{equation}  \left(\dfrac{f^{\textup{mov}}}{f}\right)_{\textup{classical}} = \dfrac{1}{1+ \dfrac{\beta x'}{\sqrt{1+x'^{2}}}} \label{Eq:Dop-class-x} \end{equation}
and for the formulas deduced from the spherical wavefront, the Doppler effects are described as functions of the coordinates of the position of the source ($ P $) and of the emission point ($ E $), by setting $ \vartheta = \theta $ or $ \theta' $ respectively.
\begin{equation}  \left(\dfrac{f^{\textup{mov}}}{f}\right)_{P} =\dfrac{\sqrt{1+x^{2}}}{\beta x+\sqrt{1-\beta^{2}+x^{2}}} \label{Eq:Dop-seen} \end{equation}
and
\begin{equation} \left(\dfrac{f^{\textup{mov}}}{f}\right)_{E} =\dfrac{1}{\sqrt{1+\beta^{2}+2 \beta \dfrac{x'}{\sqrt{1+x'^{2}}}}} \label{Eq:Dop-heard} \end{equation}

\subsection{Linear Galilean aberration}

The light Doppler effect of a star is naturally measured by pointing the telescope at this star, but we are aware that it has changed location while the light was flying towards the telescope, so that its true position is invisible. By contrast for the sound, the informations on the Doppler effect and on the location of the source are carried by different channels and can be recovered simultaneously. Indeed, the Doppler effect of the sound is naturally carried by the acoustic wave but the information on the position of the source is generally visual, i.e. carried by a light wave (Fig.1A). When the wave emitted in $ X $ arrives at the level of the receiver, the source will have continued to progress over a distance depending on the duration $ \Delta t $ of the flight of the wave from the source to the receiver. This path of length $ c \Delta t $ is the hypotenuse of a right triangle whose other two sides are the shortest distance $ D $, and the distance $ X $ separating the source from the nearest point. So Pythagoras says

\begin{subequations}
\begin{equation} (c \Delta t)^{2} = D^{2}+X^{2} \end{equation} 
from which
\begin{equation} \Delta t = \dfrac{\sqrt{D^{2}+X^{2}}}{c} = D \dfrac{\sqrt{1+x^{2}}}{c} \end{equation} 
During this time, the source will have traveled
\begin{equation} \Delta X = v \Delta t = \beta D \sqrt{1+x^{2}} \end{equation} 
or in normalized distance
\begin{equation} \Delta x =  \beta \sqrt{1+x^{2}} \label{Eq:dx} \end{equation} 
\end{subequations}

The point of emission can be calculated from the actual position of the source when the Doppler effect is detected. The angle $ \theta'= \cos^{-1} \left( -\dfrac{x}{\sqrt{1+x^{2}}}\right) $ whose origin is the point of emission, is expected to become $ \theta $ when replacing $ x $ by $  x+ \Delta x $:

\begin{equation} \begin{split} \theta & = \cos^{-1} \left( \dfrac{-(x+\beta \sqrt{1+x^{2}})}{\sqrt{1+(x+\beta \sqrt{1+x^{2}})^2}}\right) \\& \text{which can be rewritten} \\& = \cos^{-1} \left( \dfrac{\left(-\dfrac{x}{\sqrt{1+x^{2}}}\right) -\beta}{\sqrt{1+\beta^{2}-2 \beta \left(- \dfrac{x}{\sqrt{1+x^{2}}} \right)}}\right) \end{split} \label{Eq:aber-x} \end{equation}
\noindent

In this form, Eq.(\ref{Eq:aber-x}) is clearly analogous to the aberration formula Eq.(\ref{Eq:s-aber}), recovered here from a linear representation, as illustrated in Fig.5. 

\vspace{0.5cm}
\begin{center}
\includegraphics[width=8.8cm]{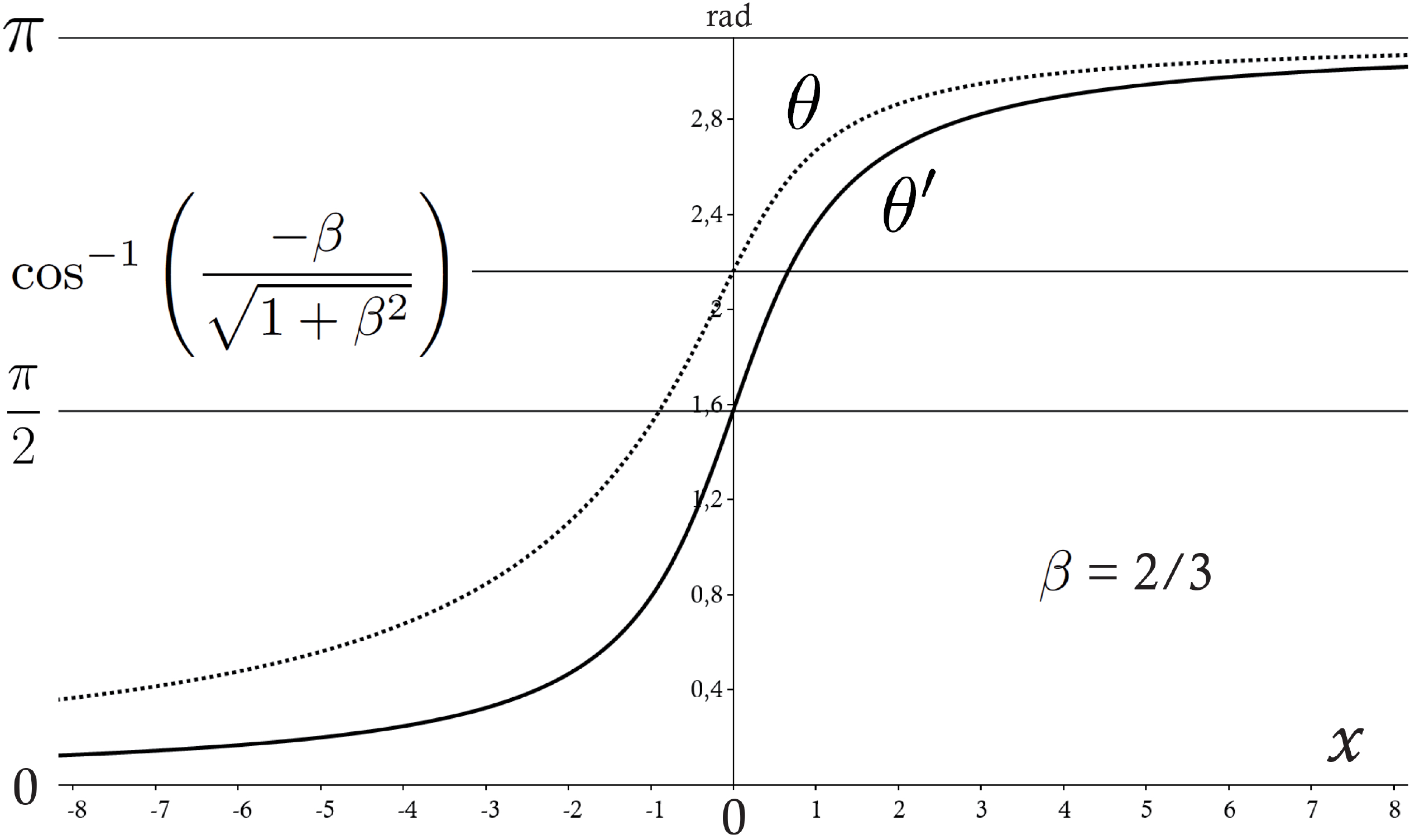}\label{fig:aber}
\end{center}
\noindent
\begin{small} \textbf{Figure 5}. Representation of the Galilean angular aberration as a function of distances. Dashed curve: angle $ \theta $ between the trajectory of the source and the direction source-receiver according to Eq.(\ref{Eq:aber-x}). Solid line curve: angle $ \theta' $ between the trajectory of the source and the line connecting the emission point to the receiver.\\ \end{small}

The introduction of Eq.(\ref{Eq:aber-x}) in the Doppler formula Eq.(\ref{eq:s-D(theta)}), indeed gives back the curve of Eq.(\ref{Eq:Dop-heard}). Conversely, the introduction in the Doppler formula Eq.(\ref{eq:s-D(theta')}) of the angle $ \theta $ obtained by conversion of $ \cos^{-1} \left( -\dfrac{x}{\sqrt{1+x^{2}}} \right) $ by the aberration formula Eq.(\ref{Eq:s-aber'}), gives the curve of Eq.(\ref{Eq:Dop-seen}). The Doppler functions derived from this approach are represented in Fig.6.

\section{The Galilean transverse Doppler effect}

The acceptance by the scientific community of the classical Doppler effect (dotted line in Fig.6), could have been favored by the mistaken intuition that the Doppler effect cancels ($ f^{\textup{mov}}/f=1 $) when the source is the closest ($ x=0 $). The absence of Doppler effect predicted by the currently admitted Eq.(\ref{Eq:dop-trad}) seems very reasonable \cite{Fowler}. It can indeed seem reasonable but it is nevertheless erroneous, as revealed by a rigorous analysis of the spherical wave (Table 1). This is typically a misleading intuition. For a moving source and static receiver, the transverse Doppler effect of sound, obtained for a reception angle $ \theta' =\pi/2 $, is

\begin{equation} \left(\dfrac{f^{\textup{mov}}}{f}\right)_{\textup{transverse}}=\dfrac{1}{\sqrt{1+\beta^{2}}} \label{Eq:gal-transverse} \end{equation} When this effect is heard, the source is at the distance $ \beta D $ from the nearest point (Table 1). By comparison, the famous relativistic transverse effect envisioned by Einstein as a possible confirmation of special relativity theory \cite{Einstein1907}, is
\begin{equation}  \left(\dfrac{f^{\textup{mov}}}{f}\right)_{\textup{transverse}}= \sqrt{1-\beta^{2}} \label{Eq:rel-transverse} \end{equation}

For small values of $ \beta $, these effects are almost identical since both are $ 1-\frac{\beta^2}{2} + \mathcal{O}(\beta^4) $ and differ only by $ \beta^4/4 $. The difference between the transverse Doppler effects calculated in the Galilean and relativistic ways, would be only 1.5\% for a speed as phenomenal as half the speed of light, making the discrimination proposed by Einstein technically very delicate. In spite of the importance of the verification of the transverse relativistic effect suggested by Einstein, only one study has confirmed this prediction \cite{Hasselkamp}. In addition, a retrospective analysis of this work raises some concerns: (i) The uncertainty margin on the result of Eq.(\ref{Eq:rel-transverse}) exceeds the value of Eq.(\ref{Eq:gal-transverse}), especially as the authors explained that they had to widen their angle to 91$ ^{\circ} $ to measure this effect. (ii) The experimental setup of these authors (see the Fig.1 of \cite{Hasselkamp}) is intriguing since the rays at 90$^{\circ}$ to the trajectory are canalized while the reception angle is diluted in a cascade of mirrors, when it is the reception angle that should be $ \pi/2 $ for the transverse effect (Fig.4 and Table 2). (iii) Finally these authors used the arithmetic mean to average wavelengths and define the so-called secondary Doppler effect (appendix B). The Galilean Doppler effect studied above for a moving source must now be completed with a series of additional subtleties for the sound.

\end{multicols}
\begin{center}
\includegraphics[width=12cm]{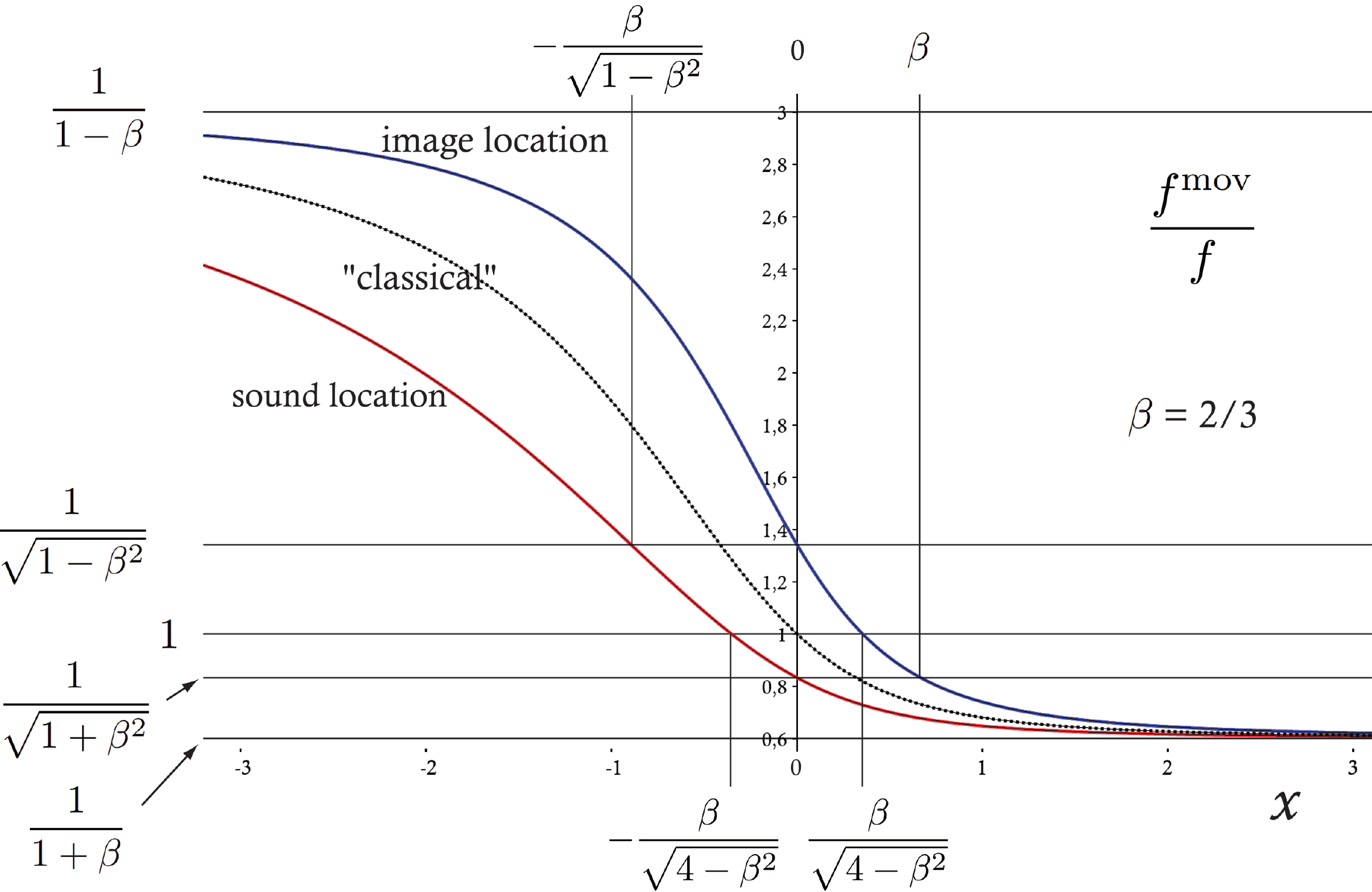}\label{fig:profils-x}
\end{center}
\begin{small}\textbf{Figure 6}. Doppler effect of the sound as a function of the relative position of a moving source on its path, for a stationary receiver in the absence of wind, expressed as a function of either the coordinate of the emission point (lower red curve drawn to Eq.(\ref{Eq:Dop-heard})), or of the position of the source detected visually (upper blue curve drawn to Eq.(\ref{Eq:Dop-seen})). The dashed curve is that of the classical Doppler formula drawn to Eq.(\ref{Eq:Dop-class-x}), shown for comparison. The increment of the $ x $ coordinate is the minimum distance between the source and the receiver. Some remarkable points from these curves are listed in the appendix A (Table 1) in comparison with those of the relativistic Doppler effect (Table 2).\\ \end{small} 
\begin{multicols}{2}

\section{The steady wind}

An additional difficulty for the sound is that its medium of propagation can itself move. The sound wave  has in fact no intrinsic material existence other than alternating compressions and depressions in the medium. A general movement of this medium, banally known as wind, is therefore expected to move the waves it contains and can of course facilitate or hinder their propagation. A first overview can be obtained for a mobile source and immobile receiver. In the same way that $ \beta$ is the ratio between the velocity of the source $ v $ and that of the wave $ c $, the ratio between the wind velocity $ v_{w} $ and that of the wave will be noted $ \omega $. The wave bubbles decentered by the velocity of the source, are now translated non-symmetrically apart from its trajectory. However, this displacement of wavefronts by the wind between stationary sources and receivers, does not induce any change in the perceived frequency, by compensation between wave speed and wavelength. Indeed, in the direction of the wind, the wave fronts progress more rapidly but their spacing lengthens in the same ratio because the wavefronts are emitted at the same rate $ f_{0} $ by the source whether there is wind or not. The successive wave crests are separated by the same time interval but by a larger spatial interval:

$$ \lambda_{w}=\dfrac{f_{0}}{c+v_{w}} $$
\vspace{0.1cm}
Hence, these wavefronts which are more spaced but travel faster arrive at the same frequency. 

\begin{center}
\includegraphics[width=7cm]{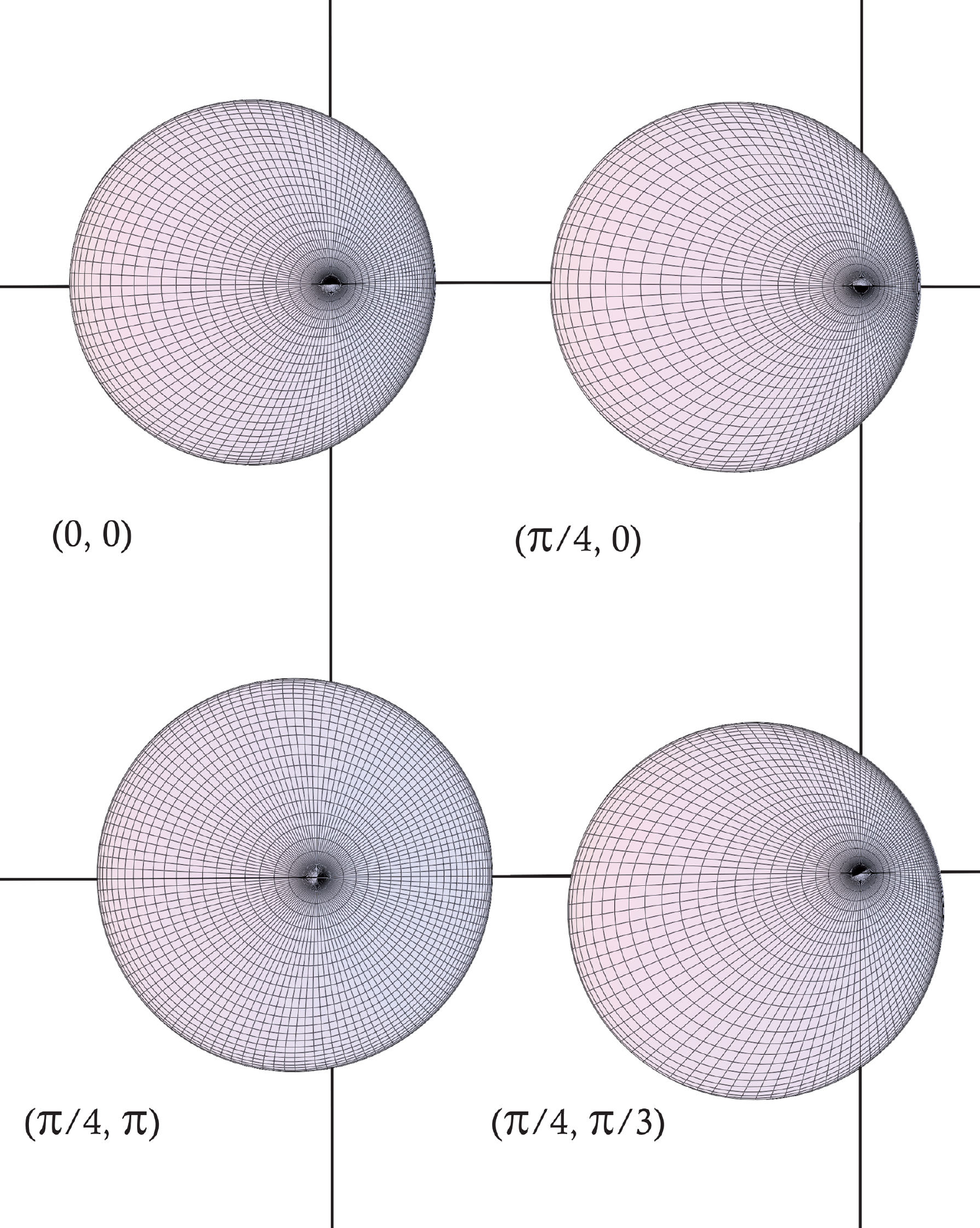}
\end{center}
\textbf{Figure 7}.\begin{small} Influence of wind on the wave bubble offset illustrated by a few pairs of values ($ \psi, \phi $), drawn to Eq.(\ref{Eq:3D-b+w}). All angular combinations maintain the spherical surface. \end{small} \\

Inversely, in the direction opposite to the wind, the wave fronts progress more slowly but their spacing is reduced in the same ratio. These wavefronts that are denser but travel more slowly also arrive at the same frequency. So in all cases for static and comobile $ S $ and $ R $, 

$$ \dfrac{c}{\lambda} = \dfrac{c_{\textup{wind}}}{\lambda_{\textup{wind}}} = f $$
\vspace{0.1cm}

This reasoning holds for all directions, that is to say for any arbitrary angle $ \theta $ in the Doppler formula. But contrary to the velocity of the wave, that of the source is considered independent of the wind, so if the source moves in the wind, the basal reference condition should be the presence of wind only. For each angle $ \theta $, the wavecrest spacing for the combined motion of the wind and the source $ \rho_{\omega+\beta} $ should be weighted by that caused by wind only $ \rho_{\omega} $, to give the Doppler effect

\begin{equation} \left(\dfrac{f^{\textup{mov}}}{f}\right)_{\textup{wind}}= \dfrac{\rho_{\omega}}{\rho_{\omega+\beta}} \end{equation}

The distances between the source and the wavefront surface depend not only on the velocity of the source compared to that of the sound ($ \beta $), but also on the velocity of the wind compared to the sound ($ \omega $), according to the Galilean law of velocity addition 
$ \beta + \omega $. These values of $ \rho $ can be calculated by repeating the approach of Cartesian to polar conversion (section 3, \cite{dmichel2022}).

\end{multicols}

Using three-dimensional polar coordinates, the sound wave bubble is shifted by the wind blowing with zenithal and asimuthal angles such that
\begin{equation} (x+\beta+\omega \sin \psi \cos \phi)^{2} + (y+\omega \sin \psi \sin \phi)^{2}+(z+\omega \cos \psi)^{2} \end{equation} 
It is converted in polar coordinates by replacing $ x $ by $\rho \sin \theta \cos \varphi $, $ y $ by $ \sin \theta \sin \varphi $ and $ y $ by $ \rho \cos \theta $, yielding the equation for $ \rho $ 
\begin{equation} \rho^{2}+2\rho \ (\beta \sin \theta \cos \varphi + \omega (\cos \theta \cos \psi + \sin \theta \sin \psi \cos (\varphi - \phi))) +\omega^{2}+\beta^{2}+2\beta \omega \sin \psi \cos \phi -1 =0 \end{equation} 
Noting
$$ A=\sin \theta \cos \varphi $$
$$ B= \cos \theta \cos \psi + \sin \theta \sin \psi \cos (\varphi - \phi) $$
$$ C = \sin \psi \cos \phi $$
the above quadratic equation reads
\begin{equation} \rho^{2}+2\rho \ (\beta A + \omega  B) +\omega^{2}+\beta^{2}+2\beta \omega C -1 =0 \end{equation} 
and its solution is
\begin{equation} \rho_{\beta+\omega} = \sqrt{1-\beta ^{2}(1 - A^{2})- \omega ^{2} (1-B^{2})-2 \beta \omega  (C - AB)}-\beta A - \omega B \end{equation}
or, retranslated in elementary components,
\begin{equation} \begin{split}  \rho_{\beta+\omega} =& \sqrt{1-\beta ^{2} \ [1 - (\sin \theta \cos \varphi)^{2}]- \omega ^{2} \ [1-(\cos \theta \cos \psi + \sin \theta \sin \psi \cos (\varphi - \phi))^{2}]} \\& \overline{-2 \beta \omega \ [\sin \psi \cos \phi - \sin \theta \cos \varphi \ (\cos \theta \cos \psi + \sin \theta \sin \psi \cos (\varphi - \phi))]}\\& -\beta \sin \theta \cos \varphi - \omega \ (\cos \theta \cos \psi + \sin \theta \sin \psi \cos (\varphi - \phi)) \end{split} \label{Eq:3D-b+w} \end{equation}
\noindent
The three-dimensional representation of this equation in Fig.7 for the angles $ \theta $ and $ \varphi $ shows that it still yields spheres. This equation in absence and presence of source motion, gives the full Doppler formula for an immobile receiver.
\begin{equation} \dfrac{\rho_{\omega}}{\rho_{\beta+\omega}} = \dfrac{\sqrt{1- \omega ^{2} (1-B^{2})}- \omega B }{\sqrt{1-\beta ^{2}(1 - A^{2})- \omega ^{2} (1-B^{2})-2 \beta  \omega (C -AB)}-\beta A - \omega B }  \end{equation}
which is developed in the appendix Eq.(\ref{Eq:sw}). In the plane $(\vec{\beta}, \vec{\omega})$, this formula reduces to

\begin{equation}\left(\dfrac{f^{\textup{mov}}}{f}\right)_{\textup{wind}}= \dfrac{\sqrt{1-\omega^2 \sin^2(\theta-\psi)} - \omega \cos(\theta-\psi)}{\sqrt{1-\left[ \omega \sin(\theta-\psi) + \beta \sin\theta \right]^{2}} - \omega \cos(\theta-\psi)-\beta \cos\theta} \label{Eq:dop-wb} \end{equation}

\noindent
of which a direct step-by-step demonstration is proposed in the appendix C, and which is illustrated in Fig.8 for the set of values: $ \beta=0.5, \omega=0.3 $ and $ \psi=\pi/4 $ or $ 3\pi/4 $.

\begin{center}
\includegraphics[width=14cm]{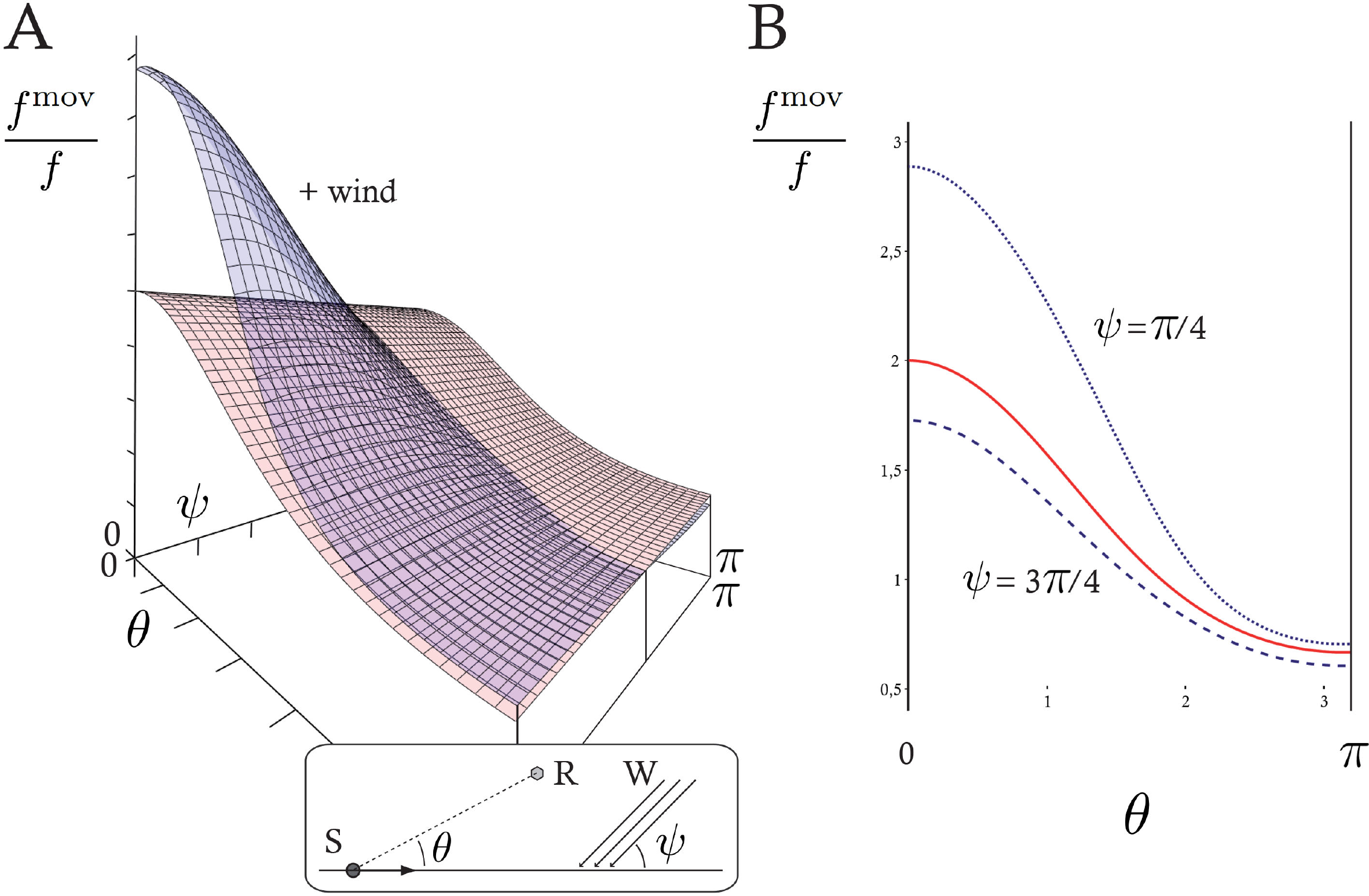}
\end{center}
\textbf{Figure 8}.\begin{small} Effect of an homogenous wind on the Doppler effect perceived by a static receiver. (\textbf{A}) Doppler effects for a mobile source and immobile detector in absence of wind (red surface) or in presence of wind blowing under the angle $ \psi $ shown in the bottom inset and ranging from 0 to $ \pi $. (\textbf{B}) Examples of Doppler profiles extracted from the panel A in the absence of wind (red continous curve), in the presence of wind blowing with the sound ($ \psi =3\pi/4 $, dashed blue curve) and against the sound ($ \psi =\pi/4 $, dotted blue curve). Curves drawn to Eq.(\ref{Eq:dop-wb}) with $ \beta=0.5 $ and $\omega=0.3 $. \end{small} \\
\begin{multicols}{2}

When the wind is collinear to the trajectory ($ \psi =\theta = 0 $), Eq.(\ref{Eq:dop-wb}) reduces to

\begin{equation} \left(\dfrac{f^{\textup{mov}}}{f}\right)_{\textup{wind}}= \dfrac{1\pm \omega}{1\pm \omega \pm \beta} = \dfrac{c\pm v_{w}}{c \pm v_{w} \pm v_{S}}\end{equation} \\ 
\noindent
which depends on the relative velocity signs (otherwise given by the cosines). A headwind ($ \omega $ negative) naturally hinders the progression of the wavefront in the $ x $ direction while a downwind ($ \omega $ positive) favors it. 

\section{Asymmetric roles of the source and receiver}

When a source $ S $ and a receiver $ R $ move one with respect to the other with a relative velocity $ v $, for the electromagnetic waves one cannot attribute the movement specifically to $ S $ or $ R $ and therefore define anything other than the relative velocity. By contrast for the sound wave, it is possible to specify the precise contributions of $ S $ and $ R $ in their relative motion, thanks to an additional point of reference offered by a sort of substratum grid. In the simple diagrams of Fig.9, this grid can be compared to the fibers of the paper of the printed article or to the pixel coordinates on the screen, which make it possible to distinguish the absolute movements of $ S $ and $ R $. Note that this spatial reference grid is not the propagation medium, because in case of wind one can also define an absolute velocity of the medium itself, relatively to some absolutely static reference grid. In ordinary experiments, the fixed spatial reference chosen is concretely the terrestrial ground assumed fixed. This major difference with light explains why for the sound, for the same relative velocity $ v $, the Doppler effect can take different values depending on whether the displacement on the grid is made by $ S $ or $ R $. 

\begin{center}
\includegraphics[width=8.5cm]{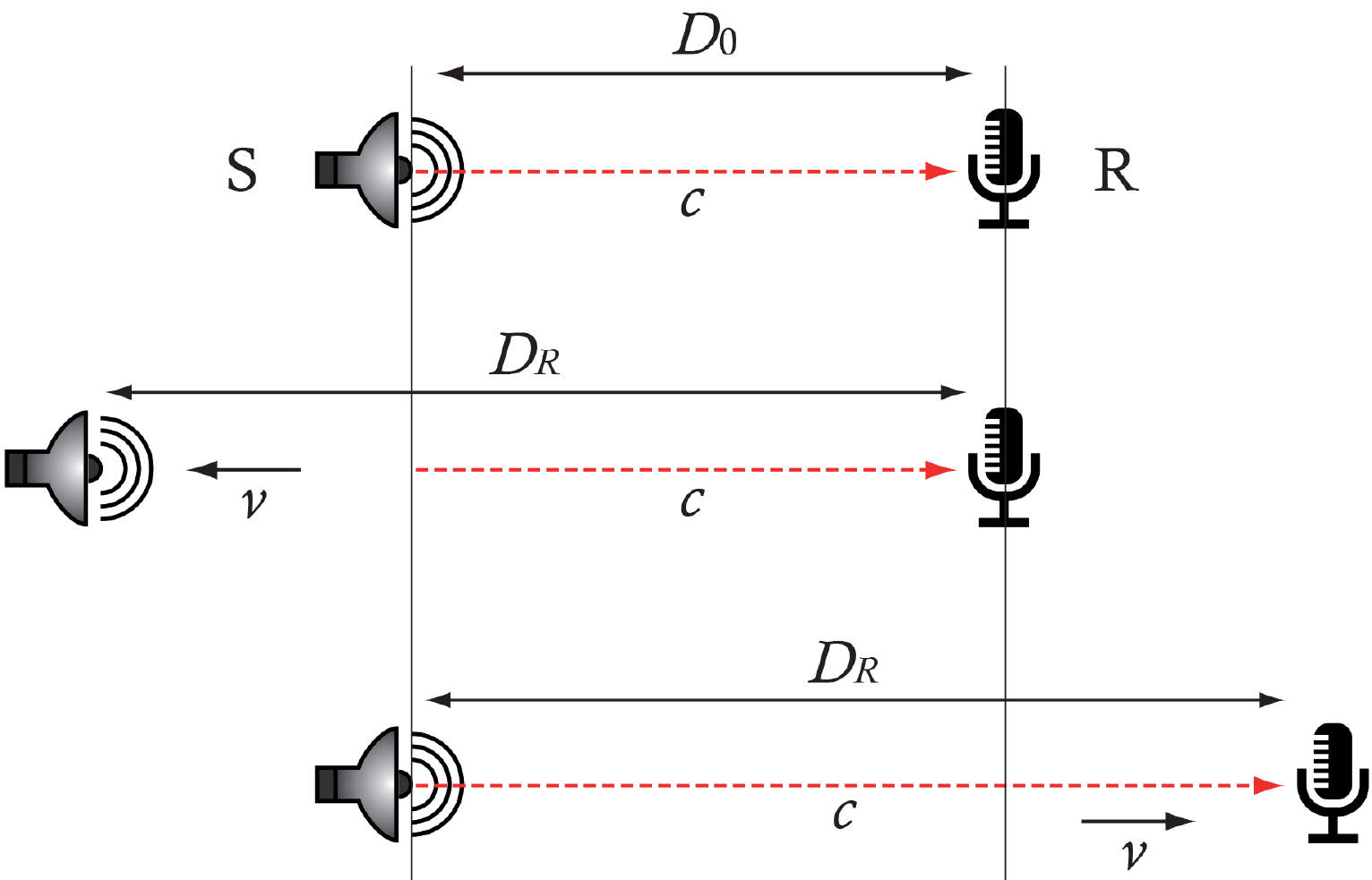}
\end{center}
\textbf{Figure 9}.\begin{small} A sound pulse is emitted at $ t_{0} $ by the source $ S $ when spaced from the receiver $ R $ by $ D_{0}$. The sound travels through a static medium relatively to which either $ R $ (middle line) or $ S $ (bottom line), is immobile. \end{small} \\

\noindent
The resolution of this asymmetry, which is simple but subtle, is an additional illustration of the fact underlined in the introduction that sophisticated mathematics are challenged in this field by simple algebra associated to good diagrams. A convenient approach in case of dual motion of the source and receiver with and without wind, is based on a principle of conservation of the wave phase \cite{Spees}. A visualization is proposed below in the case of a distancing (because it is easier to draw) between $ S $ and $ R $ (Fig.9). $ S $ and $ R $ recede form each other at relative velocity $ v $ and in addition, move relative to the static medium supporting sound propagation at speed $ c $. At time $ t_{0} $, when spaced from the receiver by $ D_{0} $, the source emits a light beam towards the receiver. Different results are expected depending on whether this is the source or the receiver which moves relatively to the static background. 

\subsection{The receiver is immobile}
When only the source moves, as shown in the middle scheme of Fig.9, its movement no longer affects the distance to travel by the sound already emitted at $ t_{0} $. Hence, the sound reaches the receiver at time $ t_{R} $ after crossing the initial distance $ D_{0} $. The duration of the sound travel is therefore
\begin{subequations}
\begin{equation}  t_{R}- t_{0}=D_{0}/c \label{Eq:gp1} \end{equation}\\
At $ t_{R} $, the new spacing between the source and the receiver has become
\begin{equation} D_{R}=D_{0} +v (t_{R}- t_{0})  \label{Eq:gp2}\end{equation}
Replacing the duration $ (t_{R}- t_{0}) $ in this equation by the value given by Eq.(\ref{Eq:gp1}), yields a distance ratio
\begin{equation} \dfrac{D_{R}}{D_{0}}=1+\dfrac{v}{c}  \label{Eq:gp3} \end{equation}
For a single period, as the source moves back, the next wavefront following that sent at $ t_{0} $ will be necessarily delayed such that 
 
 $$ \dfrac{D_{R}}{D_{0}}=\dfrac{cT_{R}}{cT_{0}} = \dfrac{f_{0}}{f_{R}} $$
\noindent
and the Doppler effect is
\begin{equation} \dfrac{f_{R}}{f_{0}}= \dfrac{1}{1+\dfrac{v}{c}} \label{Eq:gp4} \end{equation} 
\end{subequations}

\subsection{The source is immobile}
In this case (bottom situation of Fig.9), the sound travel is longer because the wavefront must reach the receiver which is concommitantly escaping.

\begin{subequations} 
\begin{equation}  t_{R}- t_{0}=D_{R}/c \label{Eq:hp1} \end{equation}
and
\begin{equation} D_{R}=D_{0} +v (t_{R}- t_{0})\label{Eq:hp2}  \end{equation}
Replacing the duration in Eq.(\ref{Eq:hp2}) by the value given by Eq.(\ref{Eq:hp1}), yields the Doppler effect
\begin{equation} \dfrac{f_{R}}{f_{0}}= 1-\dfrac{v}{c}  \label{Eq:hp3} \end{equation} 
\end{subequations}

Since the relative velocity  $ v $ can be precisely attributed to $ S $ and $ R $, it is split into $ v_{S} $ and $ v_{R} $ components. The general formula for collinear recession Doppler effect is, using velocity magnitudes,

\begin{equation} \left( \dfrac{f_{R}}{f_{0}}\right)_{r}= \dfrac{1-\dfrac{v_{R}}{c}}{1+\dfrac{v_{S}}{c}} \label{E:gp4} \end{equation} 

This simple reasoning can be repeated in the case of collinear approach

\begin{equation} \left(  \dfrac{f_{R}}{f_{0}}\right)_{a} = \dfrac{1+\dfrac{v_{R}}{c}}{1-\dfrac{v_{S}}{c}} \label{E:gp4} \end{equation} 
\noindent
and  generalized to non-collinear movements using angles. The asymmetry of the sound Doppler effect has strange consequences, long established in \cite{Joos}. For example let us consider three cases of approach where the relative velocity between $ S $ and $ R $ is the speed of sound. 

$$ c= v_{S}+v_{R} $$

If only the source moves, the Doppler effect will be $ f_{R}/f_{0}=\infty $ (sound barrier). If only the receiver advances towards the immobile source, the Doppler effect is $ f_{R}/f_{0}=2 $, and if both the source and the receiver advance one towards the other at $ c/2 $, the resulting Doppler effect is $ f_{R}/f_{0}=3 $. Such subtleties do not exist for light.

\subsection{No background medium for light}

In absence of substratum grid, it is impossible to assign the relative movement to either the source or the observer. As the contributions of $ S $ and $ R $ in the relative velocity are undistinguishable, it seems natural to average the two extreme situations described above (only $ S $ moves or only $ R $ moves). As discussed in the appendix B, the type of mathematical mean appropriate for averaging frequencies is the geometric mean. Interestingly, the geometric mean of Eq.(\ref{Eq:gp4}) and  Eq.(\ref{Eq:hp3}) is

\begin{equation} \left \langle \dfrac{f_{R}}{f_{0}} \right \rangle= \sqrt{\frac{1-\dfrac{v}{c}}{1+\dfrac{v}{c}}} \end{equation} 

In the special relativity theory, uniform motion cannot be attributed specifically to one of the relatively moving objects, so that the total distance crossed by light $ D_{L} $, is not $ D_{R} $ nor $ D_{0} $ as above, but
\begin{equation} D_{L}=c(t_{R}-t_{0})=\dfrac{c}{v}(D_{R}-D_{0}) \end{equation}

This should be the fundamental principle of distance calculation in astronomy \cite{Harrison}. As an exercise, the relationships between $ D_{R}$, $ D_{0} $, $ D_{L} $ and the resulting Doppler effect have been calculated for different theoretical modes of space expansion in \cite{dmichel2017}. But the differences between the sound and the light do not stop there.

\subsection{Triple collinear motion of the source, receiver and medium}

A wind blowing in the direction of the sound (called downwind, from left to right in Fig.9), is expected to decrease frequency whereas a headwind blowing against the sound (from right to left in Fig.9), is expected to increase frequency. The calculations can be summarized in the extreme cases of recession illustrated in Fig.9, in which either the receiver or the source remains completely immobile. The duration $ t_{R}- t_{0}$ of the sound travel, noted below $\Delta t $, depends on the velocity of the wind $ v_{w} $.

\subsubsection{Immobile receiver}
When only the source moves, as shown in the middle situation of Fig.9, 
\begin{equation} D_{R}=D_{0} +v_{S} \Delta t \label{Eq:vs} \end{equation} 
\noindent
$ \bullet $ Downwind. A wind blowing as the sound path increases it speed and push the wavefront forward, 
\begin{subequations}
\begin{equation} \Delta t = \dfrac{D_{0}}{c+v_{w}} \end{equation}
Replacing $ \Delta t $ by this value in Eq.(\ref{Eq:vs}) gives
\begin{equation}  D_{R}=D_{0} \left(1+\dfrac{v_{S}}{c+v_{w}} \right) \end{equation} 
so that
\begin{equation}  \dfrac{f_{R}}{f_{0}}= \dfrac{1}{1+\dfrac{v_{S}}{c+v_{w}}} \end{equation} 
\end{subequations}
\noindent
$ \bullet $ Headwind
\begin{subequations}
\begin{equation} \Delta t = \dfrac{D_{0}}{c-v_{w}} \end{equation}
Replacing $ \Delta t $ by this value in Eq.(\ref{Eq:vs}) gives
\begin{equation}  D_{R}=D_{0} \left(1+\dfrac{v_{S}}{c-v_{w}} \right) \end{equation} 
so that
\begin{equation}  \dfrac{f_{R}}{f_{0}}= \dfrac{1}{1+\dfrac{v_{S}}{c-v_{w}}} \end{equation} 
\end{subequations}

\subsubsection{Immobile source}
When only the source moves as in the bottom of Fig.9,
 
\begin{equation} D_{R}=D_{0} +v_{R} \Delta t \label{Eq:vr} \end{equation} 

\noindent
$ \bullet $ Downwind
\begin{subequations}
\begin{equation} \Delta t = \dfrac{D_{R}}{c+v_{w}} \end{equation}
Replacing $ \Delta t $ by this value in Eq.(\ref{Eq:vr}) gives
\begin{equation}  D_{R} \left(1-\dfrac{v_{R}}{c+v_{w}} \right) =D_{0} \end{equation} 
so that
\begin{equation}  \dfrac{f_{R}}{f_{0}}= 1-\dfrac{v_{R}}{c+v_{w}} \end{equation} 
\end{subequations}
\noindent
$ \bullet $ Headwind
\begin{subequations}
\begin{equation} \Delta t = \dfrac{D_{R}}{c-v_{w}} \end{equation}
Replacing $ \Delta t $ by this value in Eq.(\ref{Eq:vr}) gives
\begin{equation}  D_{R} \left(1-\dfrac{v_{R}}{c-v_{w}} \right) =D_{0}  \end{equation} 
so that
\begin{equation}  \dfrac{f_{R}}{f_{0}}= 1-\dfrac{v_{R}}{c-v_{w}} \end{equation} 
\end{subequations}

When both the source and the receiver are moving, the complete formulas for collinear recession are, using velocity magnitudes, \\

\noindent
$ \bullet $ Downwind
\begin{equation} \dfrac{f_{R}}{f_{0}}= \dfrac{1-\dfrac{v_{R}}{c+v_{w}}}{1+\dfrac{v_{S}}{c+v_{w}}}  \end{equation}
\noindent
$ \bullet $ Headwind
\begin{equation} \dfrac{f_{R}}{f_{0}}= \dfrac{1-\dfrac{v_{R}}{c-v_{w}}}{1+\dfrac{v_{S}}{c-v_{w}}} \end{equation}

\begin{center}
\rule{0.8\linewidth}{1pt}
\end{center}

In summary, the simplest writing, easiest to remember for all the collinear situations using velocity magnitudes, is\\

\noindent
$ \bullet $ For the recession
\begin{equation} \dfrac{f_{R}}{f_{0}}= \dfrac{c_{w}-v_{R}}{c_{w}+v_{S}} \end{equation}
with $ c_{w}=c+v_{w} $ when the wind blows with the sound and $ c_{w}=c-v_{w} $ when it blows against the sound.\\

\noindent
$ \bullet $ For the approach
\begin{equation} \dfrac{f_{R}}{f_{0}}= \dfrac{c_{w}+v_{R}}{c_{w}-v_{S}} \end{equation}
with $ c_{w}=c+v_{w} $ when the wind blows in the same direction as the sound and $ c_{w}=c-v_{w} $ when the wind is opposite to the direction of the sound.\\

The extension of these collinear equations to a general formula holding for all the combinations of relative velocities: of the source ($ \beta= v_{s}/c $), the receiver ($ \gamma = v_{R}/c $) and the medium ($ \omega=v_{w}/c $) and for arbitrary angles, would require splitting the velocity components and repeating the polar conversion approach in three-dimensional coordinates. A reduced version is obtained by confining the receiver velocity vector in the plane $ (x,y)=(\vec{\beta}, \vec{\omega}) $. Instead of staying static in the wind as in section 6, the detector moves with a velocity $\gamma = v_{R}/c $, making an angle $ \alpha $ with the source trajectory. While the propagation of the successive wavefronts results from a combination of the source and wind velocity vectors, the motion of the receiver is assumed to be autonomous and to follow a linear path whose polar equation is 
$$ \rho = \dfrac{\rho_{0}}{\sin(\alpha-\theta)} $$ crossing the successive spherical wavecrests. Depending on its direction, the receiver either comes in front of the wavecrests increasing their perceived frequency or moves away reducing the frequency. As the magnitude of the receiver velocity component towards the source is $ \gamma \cos (\alpha - \theta) $, its contribution to the Doppler effect according to \cite{Mangiarotty}, combined to those of the wind and the source defined here to take into account the aberration, gives
\end{multicols}
\begin{equation} \dfrac{f^{\textup{mov}}}{f}= \dfrac{\sqrt{1-\omega^{2} \sin^{2}(\theta-\psi)} - \omega \cos(\theta-\psi)+ \gamma \cos (\alpha - \theta)}{\sqrt{1-\left[\omega \sin(\theta-\psi) + \beta \sin\theta \right]^{2}} - \omega \cos(\theta-\psi)-\beta \cos\theta} \label{Eq:dop} \end{equation}
\begin{multicols}{2}

\noindent
of which a numerical example is shown in Fig.10. 

\label{fig:3D-srw}
\begin{center}
\includegraphics[width=7.8cm]{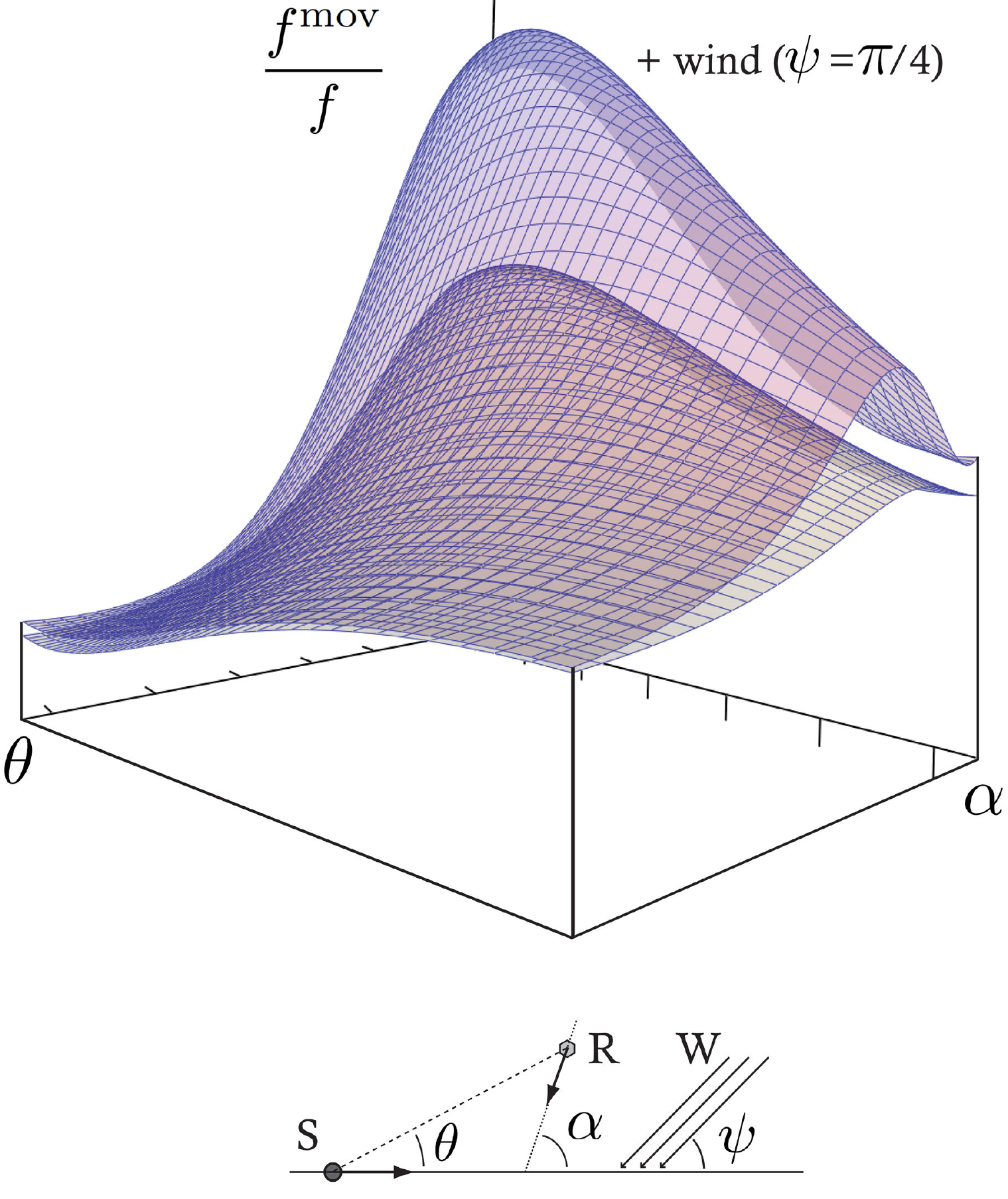}
\end{center}
\begin{small}\textbf{Figure 10}. Doppler effects drawn to Eq.(\ref{Eq:dop}) for the dual motion of the source ($ \beta=0.5 $) and of the detector ($ \gamma = 0.4 $), without wind (bottom surface) or in presence of wind ($ \omega=0.3 $) blowing with an angle $ \psi $ of $\pi/4 $ (top surface). \\ \end{small} 

This reduced formula differs in several aspects from those proposed in \cite{Young,Spees,Mangiarotty} but is not yet satisfactory and usable in practice because $\theta$ also evolves with the movement of the receiver. As illustrated in Fig.11 drawn to Eq.(\ref{Eq:dop}) simplified in the absence of wind 

$$ \dfrac{f^{\textup{mov}}}{f} = \dfrac{1+ \gamma \cos (\alpha - \theta)}{\sqrt{1-(\beta \sin\theta)^{2}} -\beta \cos\theta}$$

\noindent
the frequency perceived by the detector results from the combination of two main parameters: (i) the magnitude of its angular velocity towards the source, itself depending on the relative values of $ \alpha $ and $ \theta $, and (ii) the spacing between the successive wave crests it crosses. As these two parameters are not parallel functions of $ \theta $, this can generate internal peaks in the Doppler profiles, as illustrated in Fig.10 and Fig.11.

\label{fig:peak}
\begin{center}
\includegraphics[width=7.2cm]{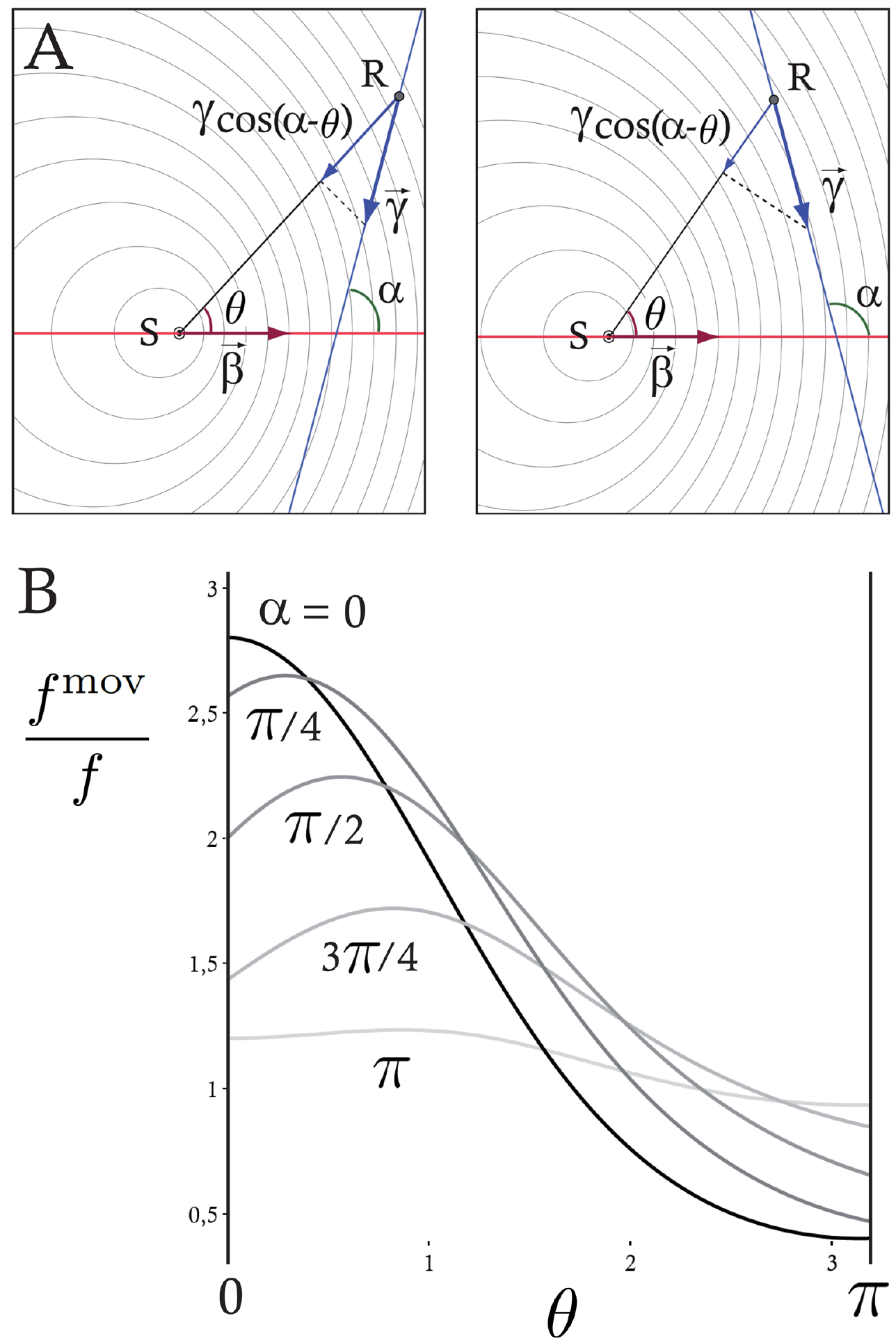}
\end{center}
\begin{small}\textbf{Figure 11}. Snapshots of source-receiver configurations (\textbf{A}) The doppler effect depends on the density of wavecrests along the line connecting $ R $ to $ S $, and on the magnitude of the revocity of $ R $ toward $ S $, which depends itself on the combination of $ \theta $ and  $ \alpha $. (\textbf{B}) Doppler profiles drawn to Eq.(\ref{Eq:dop}) in absence of wind and for different values of the angle $ \alpha $). \\ \end{small} 

In the examples shown in Fig.11A, on the one hand, an increase in $\theta$ reduces the angle $\alpha - \theta$ and therefore increases the magnitude of $ \gamma \cos (\alpha - \theta) $, but on the other hand the increase of $\theta$  also goes with a spacing of the wavecrests with inverse effects on the frequency, making Eq.(\ref{Eq:dop}) poorly useful in practice.

\section{Can wavelengths and frequencies be used indifferently to describe Doppler effects?}

Doppler effects are often described in terms of wavelength when the effects generated by waves such as the Doppler effect, sound and colors, are not determined by wavelengths $ \lambda $ but by frequencies $ f $. Frequencies and wavelengths are mutually constrained by the velocity of the wave $$ c=f\lambda $$So as long as $ c $ is constant, $ f $ and $ \lambda $ can be used indifferently to describe the colors, the Doppler effect and the sound, but one should keep in mind that the real determinant is the frequency, itself proportional for electromagnetic waves to the energy through $ E=hf $ where $ h $ is a constant (of Planck). In this case, according to the fundamental law of nature of the conservation of energy, the frequency is an invariant, so that in the absence of energy variation between reference points, any modification of $ c $ necessarily leads to a joint modification of $ \lambda $ leaving $ f $ unchanged, as are the Doppler effect and colors. However the colors are commonly associated with wavelengths (520nm for green, 630nm for red, etc), but this relation is valid only under the implicit assumption that $ c $ is constant. If the speed of light is more or less decreased, for example by passing through materials with a higher refractive index $ n $, which changes the speed of light to $ V=c/n $, where $ c $ is its speed in vacuum, the wavelengths are lengthened in the same proportion but the frequencies do not change. We have all noticed that our bathing suit does not change color when seen through water, although its wavelength has changed. The color is unchanged because it depends on the frequency only. Variations in wave velocity are even more critical for sound due in particular to the presence of wind, real or apparent, in the propagation medium (air in ordinary conditions). A particularly instructive situation to illustrate this question is shown in Fig.12. 

\subsection{Application of Fig.12 to the sound}

Two cars are driving at the same speed and in parallel, with the siren of the one shown below emitting a continuous sound. The relative speed of the two vehicles is zero, yet the wavelength received by the other vehicle is shortened by a factor of $ \sqrt{1-\beta^{2}} $ in the direction orthogonal to the travel, because the wavefronts are shifted backward by the "apparent wind" of the propagation medium. However, the driver of the upper car does not perceive any Doppler effect owing to the principle explained in section 7. The wavelength decreases by the same factor as the wave velocity, so that the frequency is unchanged, in agreement with the absence of relative motion of the two cars. We know that one swims more slowly to cross a river with a current and that the crossing period is therefore longer. It is the same in the case of Fig.12 where the role of the current is played by the apparent wind. The wavelength is the adjustment variable that guarantees the invariance of the frequency. The joint decrease of the wave speed is not only valid in the direction orthogonal to the trajectory but in all directions. Accordingly, the modification by the wind of the velocity of sound found in \cite{Spees} is

\begin{equation} c_{w} = \sqrt{c^{2}-v_{w}^{2} \sin^{2} \psi } + v_{w} \cos \psi \label{Eq:Spees} \end{equation} 

which is equivalent to the Doppler effect of Eq.(\ref{Eq:lambda-gal}) if replacing $ v_{w}/c $ by $ -\beta $. \\

\label{fig:cars}
\begin{center}
\includegraphics[width=7cm]{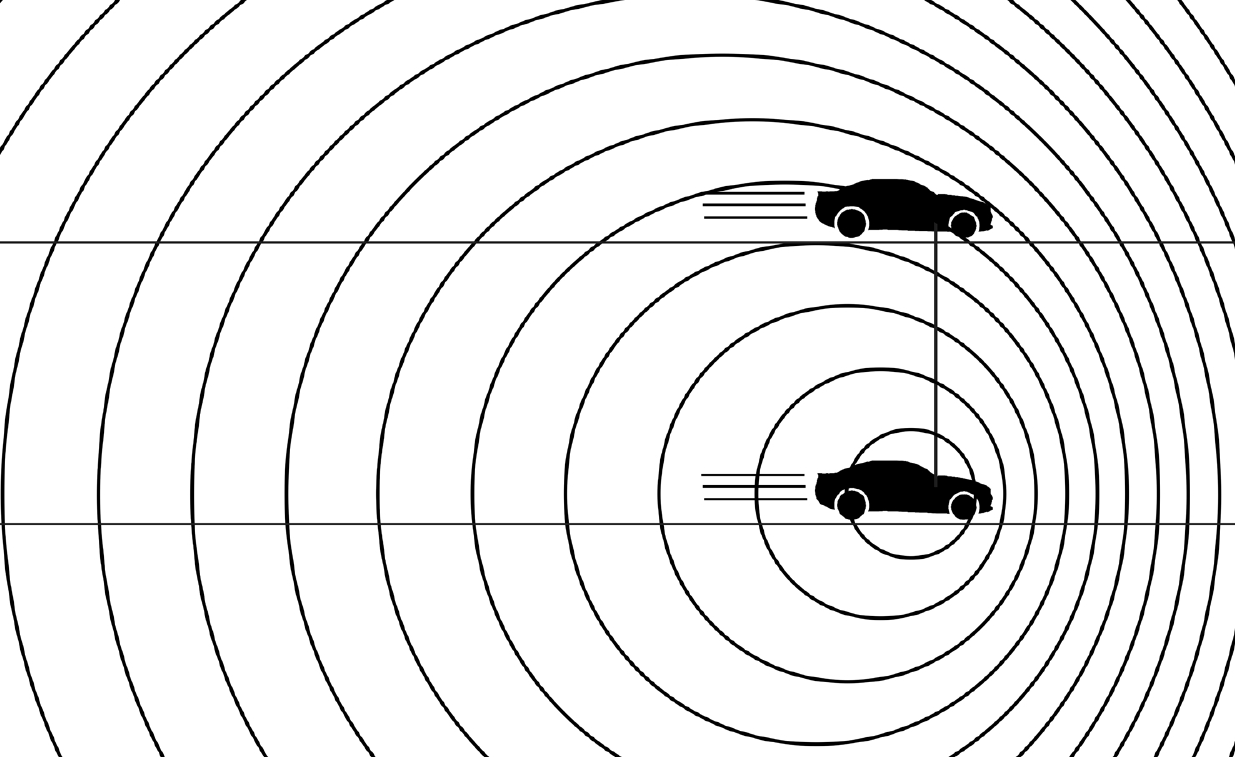}
\end{center}
\begin{small}\textbf{Figure 12}. Two cars, one of which is equipped with a siren of known frequency, drive in parallel at the same speed. The relative velocity of the two vehicles is zero and yet the wavelengths are shortened on the axis connecting them, because of the apparent wind due to the speed of the cars, which shifts the wavefronts backwards. This shortening of the wavelengths (by a factor of $ \sqrt{1-\beta^{2}} $) does not generate a Doppler effect because the velocity of sound is reduced in the same proportion, which keeps the frequency unchanged.\\ \end{small} 

\subsection{Treatment of this situation for the light}

The situation in Fig.12 is also worth considering for a light wave (by increasing the speed of luminous cars). In this case, there is no apparent wind since there is no propagation medium. For an observer considered immobile and external to the system of Fig.12, the speed of the wave $ c $ is constant but the Lorentz factor involved in time dilation will play its role by expanding the wavelengths precisely by the factor $ 1/\sqrt{1-\beta^{2}} $, thus restoring the original wavelength and maintaining unchanged both the frequency and the wavelength received by the upper driver.

\subsection{A counter-example invalidating the current sound Doppler effect}
In addition to demonstrate new formulas for the Galilean and sound Doppler effects, another approach to convince readers reluctant to question a pre-existing formulas, would be to provide a proof that this pre-existing formula is necessarily wrong as it leads to absurd results. As explained above, the driver at the top of Fig.12 does not perceive any sound Doppler effect from the siren of the bottom car; this fact is naturally expected from the absence of relative velocity between the source and the receiver. Now let us consider that the top car is immobile (in this case, of course, the diagram of Fig.12 would be a snapshot and no longer a stationary state). Which Doppler effect is then predicted by the traditional formula? (i) If the aberration effect is neglected, as it is the case in most descriptions of the classical Doppler effect, then the cosine being zero and according to the absence of transverse effect for the sound, the Doppler effect is still 1; so whether the upper car is moving or immobile, the perceived Doppler effect would be the same, which would seem very puzzling. (ii) Now if admitting that aberration also exists for a classical wave, the transverse Doppler effect would exist ($ f^{\textup{mov}}/f= 1/\sqrt{1-\beta^{2}} $) contrary to dogma. This simple example is sufficient to reveal an internal contradiction in the usual formulation of the so-called classical Doppler effect.

\section{Conclusions}
The Doppler effect is generally envisioned as a long established phenomenon corresponding to a dead branch of basic physics now confined to general education. However, a critical look reveals that the research on the acoustic Doppler effect is currently incomplete and paved with false ideas and invalid formulas. This joint study of the light and sound Doppler effects allows to perceive the specificities of each type of wave and shows that the classical Doppler effect has been strangely shaped from the knowledge of the relativistic one, while the two Doppler effects obey radically different laws. The most frequent sources of error concerning the sound Doppler effect are reviewed here. In particular, the relativistic Doppler effect is not merely a classical effect corrected by the relativistic dilatation factor. Besides, from a pedagogical point of view, we should in general be wary of the famous representations of wave crests as shifted nested circles to predict Doppler effects. Indeed, on the one hand, light wavefronts are not perceived as circles but as ellipses \cite{dmichel2022}, and on the other hand even for sound, as explained in sections 7 and 9, the nested spheres sometimes describe accurately wavelength distortions without affecting frequencies and therefore the Doppler effects either. Many more subtleties could be added to the complexity of the sound. For instance, not only does the wind modify the sounds, but it also adds its own sound. Conversely, a sound emitted by a fast source may neither be the sound of the wind nor correspond to any sound of the source at rest. The number of parameters to be taken into account to predict the Doppler effect of sound as well as the complexity of its three-dimensional formulation, show how elegant the Doppler effect of light is comparatively. For light, the absence of a propagation medium, formerly postulated under the name of ether, means that the only absolute velocity is the relative velocity \cite{Poincare1918}. The absence of the absolute substratum simplifies things considerably and is much more comfortable conceptually than the strange immaterial Galilean reference frame, in respect to which one can define absolute velocities for everything, in the case of sound for the air, the source and the detector.

\end{multicols}
\newpage
\begin{center}
\vspace{1cm}
\Huge{Appendices}
\vspace{1cm}
\end{center}
\appendix
\setcounter{equation}{0}  
\numberwithin{equation}{section}

\section{Key points to compare the Galilean and relativistic Doppler effects}

\begin{table}[!h]
\centering
\caption{Some Galilean correspondences between angles, relative distances and Doppler effects. The angle unit is the radian and the distance unit is the minimum distance between the source and the receiver. These values hold for a stationary receiver and correspond to the Doppler profiles of Fig.6. The line highlighted in gray corresponds to the transverse Doppler effect and that highlighted in yellow color is the only common point with the relativistic Doppler effect (Table 2).}
\bigskip
\label{tab:1}
\begin{tabular}{|cc|cc|c|}
\hline
& & & &  \\
\multicolumn{2}{|c|}{Origin of the angle} & \multicolumn{2}{c|}{Distance to the nearest point} & Doppler effect \\
& & & &  \\
source & point of emission & of the source (image) &  of the emission point (sound) & \\
& & & &  \\
$ \theta $  & $ \theta' $  & $ x $  &  $ x' $  & $ f^{\textup{mov}}/ f $ \\
& & &  & \\
\hline
\hline
& & &  & \\
0 & 0 &  $ -\infty $ & $-\infty $ & $ \dfrac{1}{1-\beta} $  \\
& & & &  \\
\hline
\rowcolor{LightYellow}& & & &  \\
\rowcolor{LightYellow}$ \dfrac{\pi}{2} $ & $ \cos^{-1} \beta $ & 0 & $  -\dfrac{\beta}{\sqrt{1-\beta^{2}}} $ & $ \dfrac{1}{\sqrt{1-\beta^{2}}} $  \\
\rowcolor{LightYellow}& & & &  \\
\hline
& & & & \\
$ \cos^{-1} -\dfrac{\beta}{2}  $ & $ \cos^{-1} \dfrac{\beta}{2}  $ & $ \dfrac{\beta}{\sqrt{4-\beta^{2}}} $ & $ -\dfrac{\beta}{\sqrt{4-\beta^{2}}} $ & 1  \\
& & & &  \\
\hline
\rowcolor{WhiteSmoke}& & & &  \\
\rowcolor{WhiteSmoke}$ \cos^{-1} -\dfrac{\beta}{\sqrt{1+\beta^{2}}}  $ & $ \dfrac{\pi}{2} $ & $ \beta $ & 0 & $ \dfrac{1}{\sqrt{1+\beta^{2}}} $  \\
\rowcolor{WhiteSmoke}& & & &  \\
\hline
& & & &  \\
$ \pi $ & $ \pi $ &  $ +\infty $ & $ +\infty $ & $ \dfrac{1}{1+\beta} $ \\
& & & &  \\
\hline

\end{tabular}
\end{table}

\pagebreak

\begin{table}[!h]
\centering
\caption{Some relativistic correspondences between angles, distances and Doppler effects. The line highlighted in gray corresponds to the transverse Doppler effect and that highlighted in yellow color is the only common point with the Galilean Doppler effect of Table 1.}

\label{tab:2}
\begin{tabular}{|cc|cc|c|}
\hline
& & & &  \\
\multicolumn{2}{|c|}{Origin of the angle} & \multicolumn{2}{c|}{Distance to the nearest point} & Doppler effect \\
& & & &  \\
source & emission point & of the source &  of the emission point &  \\
& & & &  \\
$ \theta $  &  $ \theta' $  & $ x $  &  $ x' $  & $ f^{\textup{mov}}/ f $ \\
& & & & \\
\hline
\hline
& & &  & \\
0 & 0 &  $ -\infty $ & $-\infty $ & $ \sqrt{\dfrac{1+\beta}{1-\beta}} $  \\
& & & &  \\
\hline
\rowcolor{LightYellow}& & & &  \\
\rowcolor{LightYellow}$ \dfrac{\pi}{2} $ & $ \cos^{-1} \beta $ & 0 & $ -\dfrac{\beta}{\sqrt{1-\beta^{2}}} $ & $ \dfrac{1}{\sqrt{1-\beta^{2}}} $  \\
\rowcolor{LightYellow}& & & &  \\
\hline
& & & & \\
$ \cos^{-1} -\dfrac{1-\sqrt{1-\beta^{2}}}{\beta}  $ & $ \cos^{-1} \dfrac{1-\sqrt{1-\beta^{2}}}{\beta}  $ & $ \dfrac{1}{\sqrt{2}}\sqrt{\dfrac{1}{\sqrt{1-\beta^{2}}}-1} $ & $-\dfrac{1}{\sqrt{2}}\sqrt{\dfrac{1}{\sqrt{1-\beta^{2}}}-1} $ & 1  \\
& & & &  \\
\hline
\rowcolor{WhiteSmoke}& & & &  \\
\rowcolor{WhiteSmoke} $ \cos^{-1} -\beta  $ & $ \dfrac{\pi}{2} $ & $ \dfrac{\beta}{\sqrt{1-\beta^{2}}} $ & 0 & $ \sqrt{1-\beta^{2}} $  \\
\rowcolor{WhiteSmoke}& & & &  \\
\hline
& & & &  \\
$ \pi $ & $ \pi $ &  $ +\infty $ & $ +\infty $ & $ \sqrt{\dfrac{1-\beta}{1+\beta}} $ \\
& & & &  \\
\hline
\end{tabular}
\end{table}
\newpage
\section{Average Doppler effects}

\begin{multicols}{2}
The concept of secondary Doppler effect supposedly specific to relativity, was further consolidated by the inappropriate use of the arithmetic mean for averaging Doppler effects. In the articles validating the relativistic Doppler effect, longitudinal \cite{Ives} and transverse \cite{Hasselkamp}, it is explained that the relativistic Doppler effects include, contrary to the classical one, a secondary transverse effect through a curious reasoning. Ives and Stilwell simultaneously measured the longitudinal wavelengths of approach ($ \lambda_{\textup{a}} $) and recession ($ \lambda_{\textup{r}} $), with and against the motion of the particles. They then compared the wavelength shifts to their so-called "center of gravity" conceived as an arithmetic mean \cite{Hasselkamp}. Knowing the relativistic longitudinal effects to be demonstrated, they calculated

\begin{equation} \label{arit-mean} \begin{split} \lambda_{\textup{mean}} & =\dfrac{\lambda_{\textup{a}}+\lambda_{\textup{r}}}{2} \\ 
& =\dfrac{1}{2}\left(\lambda_{0} \ \sqrt{\dfrac{1-\frac{v}{c}}{1+\frac{v}{c}}}+\lambda_{0} \ \sqrt{\dfrac{1+\frac{v}{c}}{1-\frac{v}{c}}}\right) \\ &=\dfrac{\lambda_{0} \ }{\sqrt{1-\frac{v^{2}}{c^{2}}}} \sim \lambda_{0} +\dfrac{\lambda_{0}}{2}\frac{v^{2}}{c^{2}} \end{split} \end{equation} 
They concluded that $ \lambda_{\textup{mean}} \neq \lambda_{0} $ due to transverse Doppler shift 
$$ \dfrac{\Delta \lambda}{\lambda_{0}} =\dfrac{\lambda_{\textup{mean}}-\lambda_{0}}{\lambda_{0}}  \sim \dfrac{1}{2}\frac{v^{2}}{c^{2}} $$

The conclusion of these authors comes from an erroneous use of the arithmetic mean. Perhaps judging the appearance of Eq.(\ref{arit-mean}) satisfactory, they did not look at what is going on for the frequency $ f_{\textup{mean}} $ corresponding to this $ \lambda_{\textup{mean}} $. Yet, since the Doppler effect during the approach for wavelengths corresponds to Doppler effect during the recession for frequencies and vice-versa, they would have found that the result is the same 

\begin{equation} f_{\textup{mean}}=\dfrac{f_{\textup{0}}}{\sqrt{1-\frac{v^{2}}{c^{2}}}} \label{Eq:fmean} \end{equation}
But for any photon, the product: frequency $ \times $ wavelength is a well known constant

\begin{equation} f \lambda=c \end{equation}
and therefore the above approach is obviously wrong as we would have

\begin{equation} f_{\textup{mean}} \ \lambda_{\textup{mean}}= \dfrac{\nu_{0} \  \lambda_{0}}{1-\frac{v^{2}}{c^{2}}} \neq c \end{equation}
In fact, the arithmetic mean used in \cite{Ives,Hasselkamp} is inappropriate for averaging Doppler effects because it cannot work for both frequencies and wavelengths. Mathematically, there are several modes of averaging which apply differently to the specific situations. These different types of averages include, when applied to two Doppler effects,
\begin{itemize}
\item The arithmetic mean: $ \dfrac{1}{2}\left (\dfrac{f^{\textup{mov}}_{1}}{f}+\dfrac{f^{\textup{mov}}_{2}}{f}  \right ) $\\
\item The geometric mean: $ \left (\dfrac{f^{\textup{mov}}_{1}}{f }\dfrac{f^{\textup{mov}}_{2}}{f}  \right )^{\frac{1}{2}} $\\
\item The harmonic mean: $ \dfrac{2f}{f^{\textup{mov}}_{1}+f^{\textup{mov}}_{2}} $\\
\end{itemize}
\noindent
The appropriate one is necessarily the geometric mean, because it is the only one holding for both periods and frequencies, such that 

\begin{equation} \left \langle f_{1},f_{2} \right \rangle=\dfrac{1}{\left \langle T_{1},T_{2} \right \rangle} \end{equation}

Besides, the use of geometric averages for wavelengths has already been applied empirically and satisfies the rule of color reflectance fusion. Therefore, let us apply the geometric mean to the Doppler effects obtained in front of and behind the closest point. We can use the linear coordinate for which the closest point corresponds to $ x=0 $, or the angular coordinate for which the closest point corresponds to $ \theta= \pi/2 $, or $ \xi= 0 $ defined such that $ \sin \xi = \cos \theta $.\\

\noindent
\textbf{For the Gallilean effect}:\\

Using $ x $ and Eq.(\ref{Eq:Dop-seen}), the geometric mean of the Gallilean Doppler effect is

\begin{equation} \forall x, \ \left \langle \dfrac{f(-x)}{f_{0}},\dfrac{f(+x)}{f_{0}} \right \rangle= \dfrac{1}{\sqrt{1-\beta^{2}}} \end{equation}

The same result is naturally obtained using angles and Eq.(\ref{eq:s-D(theta)})

\begin{equation} \forall \xi, \ \left \langle \dfrac{f(\frac{\pi}{2}-\xi)}{f_{0}},\dfrac{f(\frac{\pi}{2}+\xi)}{f_{0}} \right \rangle = \dfrac{1}{\sqrt{1-\beta^{2}}} \end{equation}
\noindent
Strikingly, this result is independant of the distance. \\

\noindent
\textbf{For the relativistic effect:}\\

Using the linear coordinate 
\begin{equation}  \forall x, \left \langle \dfrac{f(-x)}{f_{0}},\dfrac{f(+x)}{f_{0}} \right \rangle =\sqrt{\dfrac{1+x^{2}(1-\beta^{2})}{(1+x^{2})(1-\beta^{2})}} \end{equation}
and using the angular coordinate,
\begin{equation} \forall \xi, \ \left \langle \dfrac{f(\frac{\pi}{2}-\xi)}{f_{0}},\dfrac{f(\frac{\pi}{2}+\xi)}{f_{0}} \right \rangle = \sqrt{\dfrac{1-\beta^2 \sin^{2}\xi}{1-\beta^2}} \end{equation}
\noindent
which correspond to the Galilean result only for $ x=0 $ or $ \xi=0 $.
\newpage
\end{multicols}
\section{Step-by-step establishment of the 2D combination of speed and wind}
The sphere Cartesian equation
\begin{subequations}
\begin{equation} (\rho \cos \theta +\beta+ \omega \cos \psi)^2 + (\rho \sin \theta +\omega \sin \psi)^2 =1  \end{equation}
developed in polar coordinates is
\begin{equation} \rho^{2}+\omega^{2}+\beta^{2} +2\omega \rho \sin \theta \sin \psi + 2\omega \rho \cos \theta \cos \psi + 2\beta \rho \cos \theta + 2\beta \omega \cos \psi = 1 \end{equation}
and since
$$  2\omega \rho \sin \theta \sin \psi + 2\omega \rho \cos \theta \cos \psi =  2\omega \rho \cos (\theta - \psi) $$
the quadratic equation for $ \rho $ is:
\begin{equation} \rho^{2} + 2 \rho (\omega \cos (\theta - \psi)+ \beta \cos \theta)+ (\beta^{2} + \omega^{2} + 2\beta \omega \cos \psi -1) = 0\end{equation}
\end{subequations}
Its solution is
\begin{subequations}
\begin{equation} \rho= \sqrt{(\omega \cos (\theta - \psi)+ \beta \cos \theta)^{2} -\omega^{2}-\beta^{2} - 2\beta \omega \cos \psi+1}- \omega \cos (\theta - \psi)- \beta \cos \theta \end{equation}
\begin{equation} \rho= \sqrt{1+ \omega^{2} \cos^{2} (\theta - \psi) +  \beta^{2} \cos^{2} \theta + 2 \beta \omega \cos (\theta - \psi) \cos \theta -\omega^{2} -\beta^{2} - 2\beta \omega \cos \psi}- \omega \cos (\theta - \psi)- \beta \cos \theta \end{equation}
\begin{equation} \rho= \sqrt{1 -\omega^{2} + \omega^{2} \cos^{2} (\theta - \psi) -\beta^{2} +  \beta^{2} \cos^{2} \theta - 2\beta \omega \cos \psi + 2 \beta \omega \cos (\theta - \psi) \cos \theta }- \omega \cos (\theta - \psi)- \beta \cos \theta \end{equation}
\begin{equation} \rho= \sqrt{1- \omega^{2} (1- \cos^{2} (\theta - \psi)) -  \beta^{2} (1- \cos^{2} \theta) - 2\beta \omega ( \cos \psi- \cos (\theta - \psi) \cos \theta)} - \omega \cos (\theta - \psi)- \beta \cos \theta  \end{equation}
and given that: $  \cos \psi- \cos (\theta - \psi) \cos \theta = \sin \theta \sin (\theta - \psi) $,
\begin{equation} \rho= \sqrt{1- \omega^{2} \sin^{2} (\theta - \psi) -  \beta^{2} \sin^{2} \theta - 2\beta \omega \sin \theta \sin (\theta - \psi)} - \omega \cos (\theta - \psi)- \beta \cos \theta  \end{equation}
and finally
\begin{equation} \rho= \sqrt{1-\left[ \omega \sin(\theta - \psi) +  \beta \sin \theta \right]^{2} } - \omega \cos (\theta - \psi)- \beta \cos \theta  \end{equation}
\end{subequations}

This wavefront bubble shifted by the wind is the denominator of the Doppler effect, whose numerator is, in the absence of movement of the detector, the same equation reduced to the wind only ($ \beta = 0 $)

\begin{equation} \dfrac{f^{\textup{mov}}}{f} = \dfrac{\sqrt{1-\omega^2 \sin^2(\theta-\psi)} - \omega \cos(\theta-\psi)}{\sqrt{1-\left[ \omega \sin(\theta-\psi) + \beta \sin\theta \right]^{2}} - \omega \cos(\theta-\psi)-\beta \cos\theta} \end{equation}\\

With an additional dimension, the same approach gives the three-dimensional Doppler formula:
\vspace{0.1cm}

$$ \vspace{-0.4cm} \sqrt{1- \omega ^{2} \ [1-(\cos \theta \cos \psi + \sin \theta \sin \psi \cos (\varphi - \phi))^{2}]}  - \omega \ (\cos \theta \cos \psi + \sin \theta \sin \psi \cos (\varphi - \phi))$$
\begin{equation}  \dfrac{f^{\textup{mov}}}{f} = {\rule{0.8\linewidth}{1pt}} \label{Eq:sw} \end{equation}
\vspace{-0.4cm}
$$ \sqrt{1-\beta ^{2} \ [1 - (\sin \theta \cos \varphi)^{2}]- \omega ^{2} \ [1-(\cos \theta \cos \psi + \sin \theta \sin \psi \cos (\varphi - \phi))^{2}]}$$ $$ \overline{-2 \beta \omega \ [\sin \psi \cos \phi - \sin \theta \cos \varphi \ (\cos \theta \cos \psi + \sin \theta \sin \psi \cos (\varphi - \phi))]}$$ $$ -\beta \sin \theta \cos \varphi - \omega \ (\cos \theta \cos \psi + \sin \theta \sin \psi \cos (\varphi - \phi)) $$

\newpage

\section{Classical Doppler measurement}

\begin{multicols}{2}
This section aims at: (1) validating the linear Doppler approach described in this study; and (2) addressing the question of the classical transverse effect currently considered inexistent. Any ordinary film is a co-recording of image and sound, but since these two types of waves reached the camera and microphone at different speeds, they describe in fact separate moments in the recent past. To illustrate concretely this subtlety, let us analyse the shift in sound frequency during the passage of an aircraft, through the parallel analysis of image and sound.\\

\end{multicols}
\label{fig:aircraft}
\begin{center}
\includegraphics[width=14cm]{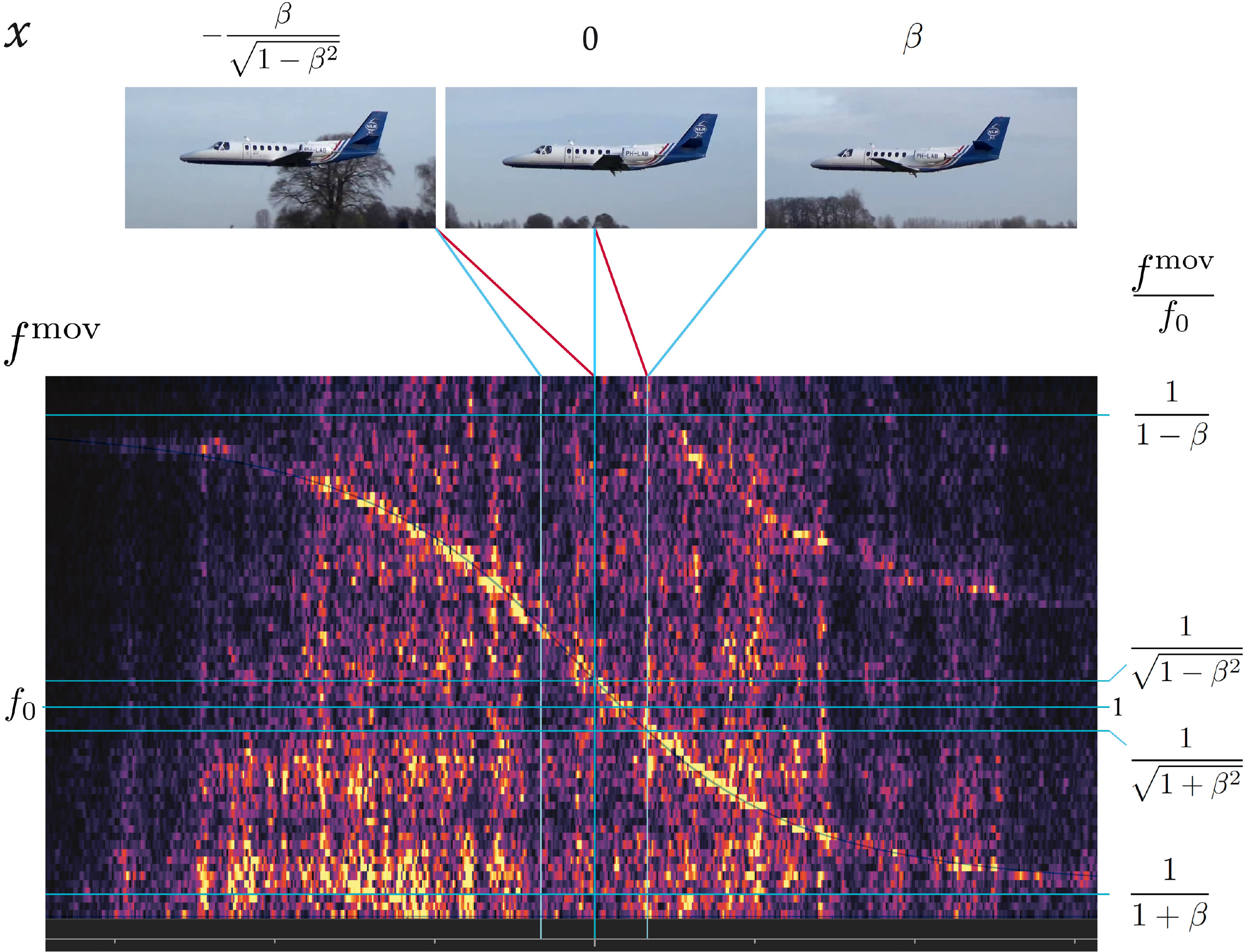}
\end{center}
\begin{small}\textbf{Figure D1}. Doppler effect illustrated by a dominant frequency recorded during the passage of an aircraft at low altitude. The blue lines connecting the images of the planes to the spectrogram indicate the actual concomitance of the sound and images on the film, while the red lines connect the recorded frequencies to their real points of emission.  \\ \end{small} 
\begin{multicols}{2}

\subsection{Determination of the aircraft speed and rest frequency}
The asympotic values of the apparent frequencies heard when the source arrives, written $ f_{a} $ and that measured when the source goes away, written $ f_{r} $, are sufficient to determine the source velocity, even in absence of knowledge of the source frequency $ f_{0} $. Indeed, $ f_{a} $ and $ f_{r} $ are related through

\begin{subequations}
\begin{equation} f_{0} = f_{a} \left (1-\beta  \right ) = f_{r} \left (1+\beta  \right )  \label{eq:beta-sound}  \end{equation} 
from which
\begin{equation} \beta = \dfrac{f_{a}-f_{r}}{f_{a}+f_{r}}\end{equation} 
\end{subequations} 

The frequencies given by the spectrogram $ f_{a} $=6750 Hz and $ f_{r} $=4338 Hz, give $ \beta = 0.2175 $ (at 15$ ^{\circ} $C, 74 m/s or 266 km/h). Once $ \beta $ is known, the equalities of Eq.(\ref{eq:beta-sound}) then allow to find the rest frequency: $ f_{0}$ = 5282 Hz. Note that although it is called rest frequency, $ f_{0} $ may not exist when the aircraft is stopped with the engines on, for instance if this sound is generated by the flux of apparent wind.

\subsection{Curve fitting and conclusions}
The theoretical equation combining the image and sound recorded simultaneously, is Eq.(\ref{Eq:Dop-seen}): $$ \dfrac{f^{\textup{mov}}}{f_{0}} =\dfrac{\sqrt{1+x^{2}}}{\beta x+\sqrt{1-\beta^{2}+x^{2}}} $$ where the ordinate is the sound Doppler effect and the abscissa $ x $ is the spatial coordinate of the source determined visually. Introducing the value of $ \beta$ measured previously in this equation gives the horizontal increment $ x=1 $. At $ x=0 $ (5156 Hz, Doppler effect of 1.025) the line of sight of the observer is perpendicular to the plane trajectory. The Doppler effect for $ x=0 $ is expected to be $$ \dfrac{f^{\textup{mov}}}{f_{0}}_{\text{orthogonal}} =\dfrac{1}{\sqrt{1-\beta^{2}}} $$
As explained in the main text, this is not the transverse Doppler effect which is

$$ \dfrac{f^{\textup{mov}}}{f_{0}}_{\text{transverse}} =\dfrac{1}{\sqrt{1+\beta^{2}}} $$

This effect (5156 Hz, Doppler effect of 0.977) is received only when the aircraft has moved away from the transverse position. Upon reception of the transverse Doppler effect, the plane it located at a distance $ \beta D $ from the closest point corresponding, given the delay of 0.323 seconds measured from the video, to 110 m from the transverse position. In summary, the accuracy of curve fitting shown in Fig.D1 can be checked by verifying the frequencies for the following two points:\\

\noindent
$ \bullet $ $ x=0 \rightarrow f^{\textup{mov}} = f_{0}/\sqrt{1-\beta^{2}} $  \\
$ \bullet $ $ x=\beta \rightarrow f^{\textup{mov}} = f_{0}/\sqrt{1+\beta^{2}} $  \\
$ \bullet $ In addition, for the agreement between the image and sound, $x=0$ must coincide with the most transverse position of the source. This can be appreciated here for instance through the apparent orientation of the wings and the alignment of the side windows of the cockpit.\\

Once these three criteria are fulfilled, the rest of the curve fits remarkably well (Fig.D1).
On Fig.D1, the blue lines join the images and the sounds which are superimposed on the video. But this apparent simultaneity is only an illusion of reception, as shown by the red lines which link the sound to the position of the plane where they were actually emitted. This offset is naturally due to the difference in speed between light and sound to get from the plane to the camera \cite{Mangiarotty}. The sound received when the airplane is seen perfectly in profile was sent at the position $ x=-\beta/\sqrt{1+\beta^{2}} $, which would belong to the curve drawn to Eq.(\ref{Eq:Dop-heard}) if added on the same diagram. 

\end{multicols}

\begin{thebibliography}{}
\begin{small}
\bibitem{Nolte2020} Nolte D.D. The fall and rise of the Doppler effect. Phys. Today 73 (2020) 30-35.
\bibitem{HRW} Halliday D., Resnick R., Walker J. Fundamentals of Physics. John Wiley \& Sons Inc. 10th ed. 2013.
\bibitem{Fowler} Fowler M. The Doppler effect.\\ https://galileo.phys.virginia.edu/classes/152.mf1i.spring02/\\DopplerEffect.htm
\bibitem{dmichel2022} Michel D. Galilean and relativistic Doppler/aberration effects deduced from spherical and ellipsoidal wavefronts respectively. Optik 250 (2022) 168242.
\bibitem{Ives} Ives H.E., Stillwell G.R., An experimental study of the rate of a moving atomic clock. J. Opt. Soc. Am. 28 (1938) 215-226. 
\bibitem{Hasselkamp} Hasselkamp D., Mondry E., Scharmann A. Direct observation of the transversal Doppler-shift. Z. Phys. A 289 (1979) 151-155.
\bibitem{Poincare1918} Poincar\'e H. Le principe de Relativit\'e. La dynamique de l'\'electron. Revue G\'en\'erale des Sciences Pures et Appliqu\'ees. 19 (1918) 386-402.
\bibitem{Poincare1905} Poincar\'e H. Sur la dynamique de l'\'electron. C.R. Acad. Sci. 140 (1905) 1504-1508.
\bibitem{Einstein} Einstein A. Zur Elektrodynamik bewegter K\"orper (On the electrodynamics of moving bodies), Annal. Phys. 17 (1905) 891-921.
\bibitem{Joos} Joos G., Freeman, I.M. Theoretical Physics. Hafner publishing company NY. 1958.
\bibitem{Mangiarotty} Mangiarotty R.A., Turner B.A. Wave radiation doppler effect correction for motion of a source, observer and the surrounding medium. J. Sound Vib. 6 (1967) 110-116.
\bibitem{Einstein1907} Einstein A. Possibility of a new examination of the relativity principle, Annal. Phys. 23 (1907) 197-198.
\bibitem{Spees} Spees A.H. Acoustic doppler effect and phase invariance. Am. J. Phys. 24 (1956) 7-10.
\bibitem{Harrison} Harrison E. The redshift-distance and velocity-distance laws. Astrophys. J. 403 (1993) 28-31.
\bibitem{dmichel2017} Michel D. Analytical relationships between source-receiver distances, redshifts and luminosity distances under pure modes of expansion. Adv. Astrophys. 2 (2017) 217-230.
\bibitem{Young} Young R.W. The Doppler effect for sound in a moving medium. J. Acoust. Soc. Am. 6 (1934) 112-114.
\end{small}
\end{thebibliography}
\end{document}